# The impact of physicochemical features of carbon electrodes on the capacitive performance of supercapacitors: A machine learning approach

Sachit Mishra[a,b,1], Rajat Srivastava[a,c,1], Atta Muhammad[a,d], Amit Amit[a], Eliodoro Chiavazzo[a], Matteo Fasano[a,*], and Pietro Asinari[a,e]

[a] Department of Energy "Galileo Ferraris", Politecnico di Torino, Corso Duca degli Abruzzi 24, 10129 Torino, Italy.
[b] IMDEA Network Institute, Universidad Carlos III de Madrid, Avda del Mar Mediterraneo 22, 28918 Madrid, Spain.
[c] Department of Engineering for Innovation, University of Salento, Piazza Tancredi 7, 73100 Lecce, Italy.
[d] Department of Mechanical Engineering, Mehran University of Engineering and Technology, SZAB Campus, Khairpur Mir's 66020, Sindh, Pakistan.
[e] Istituto Nazionale di Ricerca Metrologica, Strada delle Cacce 91, 10135 Torino, Italy.
* Corresponding author.
Email address: matteo.fasano@polito.it (Matteo Fasano).
[1] Equal contributors.

**Abstract**
Hybrid electric vehicles and portable electronic systems use supercapacitors for energy storage owing to their fast charging/discharging rates, long life cycle, and low maintenance. Specific capacitance is regarded as one of the most important performance-related characteristics of a supercapacitor's electrode. In the current study, Machine Learning (ML) algorithms were used to determine the impact of various physicochemical properties of carbon-based materials on the capacitive performance of electric double-layer capacitors. Published experimental datasets from 147 references (4899 data entries) were extracted and then used to train and test the ML models, to determine the relative importance of electrode material features on specific capacitance. These features include current density, pore volume, pore size, presence of defects, potential window, specific surface area, oxygen, and nitrogen content of the carbon-based electrode material. Additionally, categorical variables as the testing method, electrolyte, and carbon structure of the electrodes are considered as well. Among five applied regression models, an extreme gradient boosting model was found to best correlate those features with the capacitive performance, highlighting that the specific surface area, the presence of nitrogen doping, and the potential window are the most significant descriptors for the specific capacitance. These findings are summarized in a modular and open-source application for estimating the capacitance of supercapacitors given, as only inputs, the features of their carbon-based electrodes, the electrolyte and testing method. In perspective, this work introduces a new wide dataset of carbon electrodes for supercapacitors extracted from the experimental literature, also giving an instance of how electrochemical technology can benefit from ML models.

# 1. Introduction

Electrochemical capacitors (supercapacitors) are electrochemical devices that are extensively used for energy storage due to promising characteristics such as high-power density, electrochemical stability, fast charge/discharge rates, safe operation mode, high power density, and long cycle life. [1-3] These characteristics enable their use in a broad range of energy storage applications, e.g., for hybrid electric vehicles, portable electronics, and memory backup systems. [4-6] Other than energy storage, there are some other interesting applications of supercapacitors such as heat-to-current conversion of low-grade thermal energy [7] and renewable energy extraction using a supercapacitor from water solutions. [8] Supercapacitors have been primarily classified into two types based on their charge storage mechanism: (i) electric double-layer capacitors (EDLCs), which store electrical charge via ion adsorption at the electrode surface, and (ii) pseudo-capacitors, which store charges via reversible Faradaic redox reactions (see Figure 1). Generally, EDLCs have superior cycle stability but lower specific capacitance in comparison to pseudo-capacitors, which have a high specific capacitance but a low power density and poor cycle stability instead. [9,10] The current study focuses on EDLC supercapacitors and their optimization.

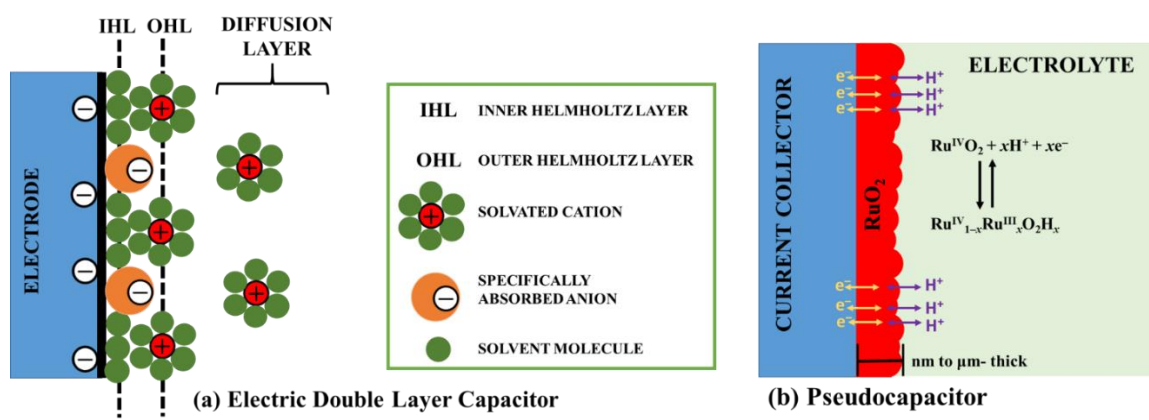

Figure 1. Schematic diagram of supercapacitors: (a) Electric double-layer capacitor (EDLC); (b) example of pseudocapacitor based on ruthenium oxide.

For the optimization of the electrochemical performance of EDLC supercapacitors, it is critical that the electrode materials have commendable physicochemical properties, including appropriate pore size distribution, high specific surface area, high electrical conductivity, as well as electrochemical and mechanical stability for good cycling performance. [9] Numerous materials have been synthesised and used as supercapacitor electrodes in recent years, including porous carbon, [11-13] hierarchical porous carbon, [14-16] activated carbon, [10,17,18] graphene, [9,19,20] rGO-PANI nanocomposite, [21] carbon nanotubes, [22-24] $In_2O_3$-loaded porous carbon, [25] and carbon aerogels. [26,27] Among these electrode materials, carbon is the most frequently used due to its versatility and uniqueness. [28] It exists in various forms (e.g., graphite, diamond), dimensionalities (fibres, fabrics, foams, and composites), ordered and disordered structures (depending on the degree of graphitization), with commendable electrical conductivity. [29,30] The catalytic, optical, mechanical, and electrochemical properties of carbon make it an excellent material for energy conversion and storage applications. [31] Additionally, its well-established synthesis and activation methods enable its use as an electrode in supercapacitors with an appropriate pore size distribution. [32]

Porous and activated porous carbon (AC) and hierarchical porous carbon (HPC) have been proposed in the literature for carbon electrodes (see Supplementary Note 1 for a detailed review). Because of its high specific surface area (SSA), improved electrical conductivity, adjustable pore sizes, electrochemical stability and low cost, porous carbon offers significant promise for use as the electrode. [33-36] These properties make AC an excellent material for a variety of applications, including water purification, gas separation and storage, and electrode materials for capacitors, fuel cells and batteries. [37] Differently from AC, hierarchical porous carbon material contains pores in a wide range of length scales, namely macro- (>50 nm), meso- (2–50 nm), and micro- (<2 nm) scales. The presence of macropores in HPC allows high-rate ion transport and acts as an ion reservoir. Furthermore, the interconnected mesopores provide low resistance pathways for the diffusion of ions; whereas the high SSA of micropores enhances the adsorption of ions at the pore surface. [38] These unique properties of HPC gained recent interest in the selection of electrode materials for supercapacitors.

Besides the SSA and pore volume, there are also several other factors that influence performance of electrodes in supercapacitors, such as surface functional groups and conductivity. These can be modified by introducing heteroatoms – HA (nitrogen, oxygen, sulphur, etc.) in the carbon electrodes, which do not only enhance the wettability, but also improve electronic conductivity of activated carbon. [39] Nitrogen doping on carbonaceous material as electron donor is useful for enhancing the specific capacitance via faradaic reaction and enhancing wettability. [16] Similarly, oxygen doping improves the surface wettability, which in turn improves the supercapacitor performance. [39] Sulphur doping on carbonaceous material increases its bandgap, thus enhancing the electron donor properties and changing the electronic density of state. Sulphur doping also increases wettability, which in turn decreases the diffusion resistance that occurs between the electrode and electrolyte ions. [40]

In perspective, graphene is a promising electrode material for supercapacitors too, due to a high electrical conductivity, high SSA, and excellent mechanical strength. [41,42] Its porous structure also facilitates charge transport in the supercapacitor. The SSA of graphene is highly tuneable according to the requirement of supercapacitor electrode for energy storage applications. Also, the presence of highly movable free $\pi$ electrons on its orbital are responsible for the exceptionally high electrical conductivity. [41] Furthermore, the electrical behaviour of graphene can be improved through functionalization [43] and heteroatom doping. [44]

Numerous attempts have been made to increase the specific capacitance of supercapacitors by utilizing different types of carbon electrodes with varied pore size distributions, high specific surface area, diverse morphologies, and modified surface chemistry. [45] However, the influence of these physicochemical parameters on the specific capacitance of supercapacitors has not been completely understood. Additionally, conventional theories and models are incapable of capturing with sufficient accuracy the microscopic details of the underlying physical mechanisms affecting ion transport, which are essential for accurately predicting the capacitive performance of supercapacitors. Recent advances in machine learning (ML) algorithms and their application to physics-based systems have made it possible to recognize the effects of various physicochemical features of carbon-based electrode materials in enhancing the specific capacitance of supercapacitors. In detail, Zhu *et al.* [46] used artificial neural network (ANN) algorithm to predict the specific capacitance of carbon-based supercapacitors. They collected 681 data entries from the published experimental papers, with information about specific surface area, pore size, presence of defects, nitrogen doping level, and potential window. The authors concluded that ANN yields better predictability of specific capacitance than linear regression and Lasso methods. However, the ANN method could not discriminate the impact of each feature separately. Instead, Su *et al.* [47] interpolated the specific capacitance of carbon-based electric double layer capacitors using four different ML models, namely linear regression (LR), support vector regression (SVR), multilayer perception (MLP), and regression tree (RT). The authors ranked the performance of the different ML models as follows: RT > MLP > SVR > LR. They found out that the specific surface area, potential window, and heteroatom doping enhance the specific capacitance of EDLC supercapacitors. Nevertheless, the authors did not analyse the effect of pore volume and size of electrodes. Finally, Zhou et al. [48] proposed a ML model to determine the features with stronger impact on the specific capacitance and power density of supercapacitors, limiting their analysis only to activated carbon materials for the electrodes and a 6 M KOH electrolyte.

Although some data-driven analyses of the relation between a few features of supercapacitors and their specific capacitance have been reported in previous studies, a comprehensive study on more physicochemical features, electrode materials, methods of testing and electrolytes has been hindered by the limited number of entries in the considered database. In this work, we first created a larger dataset by extracting data from 147 experimental research articles on supercapacitors comprising carbon-based electrodes. The resulting curated dataset is made of 4899 entries and primarily contains information about the specific surface area, the presence of defects, the pore volume and size of pores, the potential window, the current density as well as the nitrogen and oxygen content of the carbon-based electrode materials. Additionally, the importance of categorical variable such as testing method, electrolyte, and carbon structure of the electrode on the specific capacitance was studied for the first time. ML algorithms were then applied to this dataset to identify those characteristics of the electrode material that significantly affect their capacitive performance, and to develop the best model possible for predicting the specific capacitance of supercapacitors. To ease the transferability of results, we developed SUPERCAPs, an open-source software to estimate the specific capacitance of carbon-based EDLC according to the structural features of electrodes, the electrolyte solution and method of testing.

## 2. Materials and methods

### 2.1. Dataset creation

To develop the dataset, we extracted information from 147 research articles on carbon-based electrode supercapacitors, collecting 4899 data entries (see Supplementary Dataset and the Supplementary Note 2 for a detailed list of data sources). Each data entry includes information related to carbon electrodes (*i.e.*, pore size, pore volume, etc.), the test system (*i.e.*, electrolyte, potential window, current densities), and the resulting specific capacitance. The latter is defined as $C = \frac{\varepsilon S}{d}$, where, $C$, $\varepsilon$, $S$, and $d$, are the specific capacitance, permittivity of electrolyte, surface area of electrode-electrolyte interface, and charge separation distance, respectively.

The various parameters included in the dataset that characterize the electrodes and the test system are as follows:

1. *Specific surface area (SSA, [m$^2$/g])*. The specific capacitance of EDLC supercapacitors depends on the adsorption of electrolyte ions on the electrode surface and directly depends on the surface area of the electrode material. Thus, to enhance the specific capacitance, a high specific surface area of electrode material is preferable. [1,29]
2. *Pore size (PS, [nm])*. The presence of micro/mesopores in carbon-based electrodes provides efficient pathways for the electrolyte ions transports, which leads to rapid ionic diffusion in the supercapacitor. [49-51]

3. *Pore volume (PV [cm³/g])*. This feature is related to PS, with an additional normalization with respect to the mass of the electrode.
4. *Ratio between D and G peaks ($I_D/I_G$, [-])*. The high ratio of intensities between peaks D and G represents the increase in defects, which leads to a decrease in the electrical conductivity of carbon-based electrodes. The decrement in the electrical conductivity of electrode material affects the capacitive performance of the supercapacitor. [49]
5. *Nitrogen content in the electrode (N%, [%])*. The nitrogen doping in the carbon matrix electrode material improves the specific capacitance by Faradaic reaction. It does not only enhance the charge mobility on carbon surfaces, but it also increases its wettability. [16,52]
6. *Oxygen content in the electrode (O%, [%])*. The oxygen content in the electrode material improves the wettability of the electrode surface in the electrolyte, which enhances the electrochemical performance of the supercapacitor. [39]
7. *Sulphur content in the electrode (S%, [%])*. Sulphur is most reactive element among heteroatom doping elements due to its unpaired electrons and wider bandgap. Increased specific capacitance results from the sequence of Faradaic reactions on sulphur doped carbonaceous materials. [40]
8. *Potential window (PW, [V])*. Potential window is a range of potentials in which no Faradaic reaction occurs, implying that material and electrolyte are stable when the potential is applied in this range. It is dependent on the type of material and electrolyte.
9. *Current density (I, [A/g])*. In porous carbonaceous electrodes, ions of the electrolyte do not have sufficient time to reach the microporous surface of the electrode at a high current density due to a fast-charging rate. Therefore, increasing current density degrades the capacitive performance of the supercapacitor. [53,54]

Furthermore, the following categorical variables have been also included in the dataset:
1. *Electrolyte type.* The type of electrolyte is crucial to supercapacitor performance. A good electrolyte has a broad potential window, strong electrochemical stability, high ionic concentration, and conductivity. Electrolytes are classified into three types: aqueous, organic liquid and ionic liquid. [55]
2. *Method of testing specific capacitance.* Two-electrode and three-electrode method are the two methods for evaluating the specific capacitance. The two-electrode method consists of working and counter electrodes, where the potential is supplied, and the resultant current is obtained at either working or counter electrode. The three-electrode system consists of working electrode, counter electrode, and reference electrode. The reference electrode serves as a reference for measuring and adjusting the working electrode potential, without transmitting any current.
3. *Electrode structure.* AC, HPC, and heteroatom (HA)-doped electrodes have a significant effect on their performance, as comprehensively discussed in the Supplementary Note 1.

Figure 2 (a)-(g) shows the influence of the various physicochemical parameters on the specific capacitance of supercapacitors at a current density of 1 A/g for the whole dataset (note that, due to lack of data at 1 A/g, the relationship between specific capacitance and sulphur doping percentage is not presented); whereas Figure 2 (h) shows the relationship between specific capacitance and different current densities. While SSA shows a certain correlation with $C$ in agreement with previous results,

[1,29] the other features of carbon-based supercapacitors have a less clear and nonlinear influence on the specific capacitance, therefore requiring advanced data analysis tools to fully understand it.

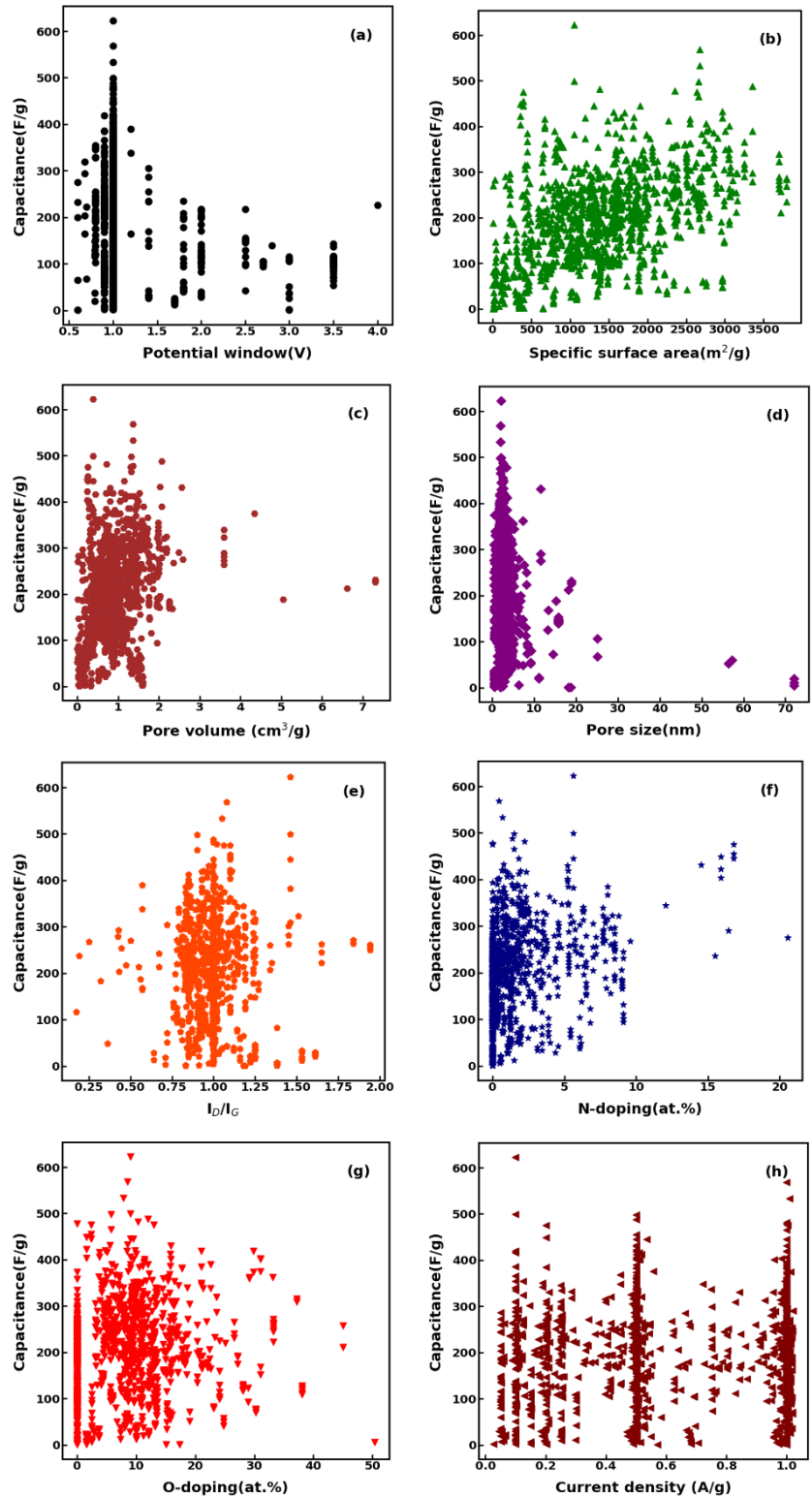

Figure 2: Relation between the specific capacitance of supercapacitors in the curated dataset and (a) potential window, (b) specific surface area, (c) pore volume, (d) pore size, (e) $I_D/I_G$, (f) N-doping (wt. %), and (g) O-doping (wt. %) at current density of I = 1 A/g. (h) Relationship between specific capacitance and different current densities.

Prior to applying the regression algorithms to the dataset containing the physicochemical properties and specific capacitance of carbon-based electrodes, it is necessary to pre-process the data to remove possible gaps and outliers. The process of improving data quality is known as data curation, and it entails the following activities: [56]

- *Data integration.* The raw data entries in the dataset were derived from various research articles that use a variety of physical units to represent parameters (for example, SSA can be expressed in $m^2/g$ or in $cm^2/g$). We maintained consistent physical units across the dataset and converted them whenever needed.
- *Outlier detection.* The dataset was analysed to identify missing values, erroneous values extracted from research articles, or values that are incorrectly formatted, which could skew the results.

Once data curation is completed, the clean dataset (4538 data entries) can be used to train and test the regression algorithms.

## 2.2. Regression model and metrics

Five approaches were adopted to carry out the regression of the target specific capacitance from the physicochemical features of the supercapacitors (see Supplementary Note 3 for details), namely the Ordinary Least Square Regression (OLS) method and four ML approaches: Support Vector Regression (SVR); Regression Decision Tree (DT); Random Forest Regression (RF); Extreme Gradient Boosting Regression (XGBoost).

OLS is one of the most common regression models, where the unknown parameters of linear regression are estimated by lessening the sum of the squares of the differences between the target responses of the sample data and the value foreseen by a linear function of explanatory variables. [57]

SVR is a well-established supervised machine learning approach for predicting discrete values. SVR operates on the same principle as Support Vector Machine (SVM). The primary principle of SVR is to determine the best fit line. Support vectors are the results of ideal hyperplanes, which classify unseen datasets that support hyperplanes. [58] SVM defines an optimal hyperplane as a discriminative classifier, whereas – in SVR – the best fit line is the hyperplane with the most point. The hyperplane in a two-dimensional region is a line separating into two segments wherein each segment is placed on either side. For instance, multiple line data classification can be done with two distinct datasets (*i.e.,* green and red) and used to propose an affirmative interpretation. However, selecting an optimal hyperplane is not an easy job, as it should not be noise sensitive, and the generalization of datasets should be accurate. [59] Pertinently, SVM is used to determine the optimized hyperplane that provides considerable minimum distance to the trained dataset. [58,59] SVR attempts to minimize the difference between the real and predicted values by fitting the best line under a certain threshold value. The distance between the hyperplane and the boundary line is the threshold value. [60]

DT constructs the regression or classification models based on the data features in the tree's configuration. In a tree, every node is related to the property of a data feature. Moreover, it either predict the target value (regression) or predict the target class (classification). The closer the nodes in a tree are, the greater their influence. [61] Some benefits of the DT include the capability of handling both categorical and numerical data.

RF is an ensemble learning technique that can perform both regression and classification tasks utilizing the multiple decision trees. During training, the algorithm generates a large number of decision trees using a probabilistic scheme; [62] every tree is trained on a bootstrapped sample of the original training data and finds a randomly selected subset of the input variables to determine a split (for each node). Every tree in the RF makes its own individual prediction or casts a unit vote for the most popular class at input *x*. These predictions are then averaged in case of regression or the majority vote determines the output in case of classification. [62] The core concept is to use numerous decision trees to determine the final output rather than depending on individual decision trees.

XGBoost is one of applications of gradient boosting machines mainly designed for speed and performance in supervised learning. In supervised learning, various features in the training data are utilized to predict the target values. XGBoost applies the tree algorithms to a known dataset and categorises the data accordingly. [63] In this model, decision trees are constructed sequentially. Weights are very significant in XGBoost: they are assigned to all the independent variables which are then input into the decision tree which determines the outcomes. The weight of variables predicted wrong by the tree is increased and these variables are fed to the second decision tree. These distinct classifiers are then combined to form an efficient and precise model. XGBoost can be used for both classification and regression problems. [64]

To predict the performance of the regression models, the $n$ predicted results ($y_i$) were compared to the original ones ($\hat{y}_i$) using the following metrics: [65]

- *Root mean square error (RMSE):*

$$\text{RMSE} = \sqrt{\frac{1}{n}\sum_{i=1}^{n}(y_i - \hat{y}_i)^2} . \tag{1}$$

A RMSE value closer to zero denotes a better prediction.

- *Coefficient of determination ($R^2$):*

$$R^2 = 1 - \frac{\sum(y_i - \hat{y}_i)^2}{\sum(y_i - \bar{y})^2}, \tag{2}$$

where $\bar{y}$ is the mean of $\hat{y}_i$ values. An $R^2$ value closer to one represents better prediction.

- *Bias factor ($b'$):*

$$b' = \frac{1}{n}\sum_{i=1}^{n}\frac{y_i}{\hat{y}_i}. \tag{3}$$

The value predicted by the model is unbiased if $b' = 1$.

- *Mean absolute percentage error (MAPE):*

$$MAPE = \frac{100\%}{n}\sum_{i=1}^{n}\left|\frac{y_i - \hat{y}_i}{y_i}\right|. \tag{3}$$

A MAPE value closer to zero denotes a better prediction.

## 3. Results and discussion

### 3.1. Correlation analysis

The correlation analysis is a statistical technique used for determining the strength of a relationship between a pair of parameters (variables). [66] To estimate the correlations between each pair of parameters, the Spearman's rank correlation coefficient ($r_s$) was used, in order to encompass nonlinear relations as well. [67] The correlation (absolute values of $r_s$) between the various supercapacitor's parameters (*i.e.*, possible descriptors) and their specific capacitance (*i.e.*, figure of merit) at different intervals are presented in Table 1. This dataset analysis revealed that the specific capacitance of the supercapacitor had a moderate correlation with the SSA, a weak correlation with the nitrogen and oxygen content of the carbon electrode, and a very low or negligible correlation with the remaining parameters. These findings suggest that SSA, N%, and O% are important parameters for the enhancement of the capacitive performance of supercapacitors. In addition, the cross-correlation analysis between all considered parameters depicted in Figure 3 shows that the physicochemical parameters of carbon electrodes are largely independent of one another, except for SSA, PV, and PS, which have a weak or a moderate geometrical intercorrelation. As a result, we can assert that the physicochemical parameters of the carbon electrode have mostly an independent effect on the supercapacitor's specific capacitance, thus they should be better considered separately from each other.

| $|r_s|$ | Level | Parameters |
|---|---|---|
| 0.0 – 0.19 | very weak | C vs PW, C vs PV, C vs PS, C vs $I_D/I_G$, C vs I |
| 0.2 – 0.39 | weak | C vs N%; C vs O% |
| 0.4 – 0.59 | moderate | C vs SSA |
| 0.6 – 0.79 | strong | - |
| 0.8 – 1.00 | very strong | - |

Table 1: Spearman's rank correlation coefficient between the specific capacitance and different features of carbon supercapacitors in the considered dataset.

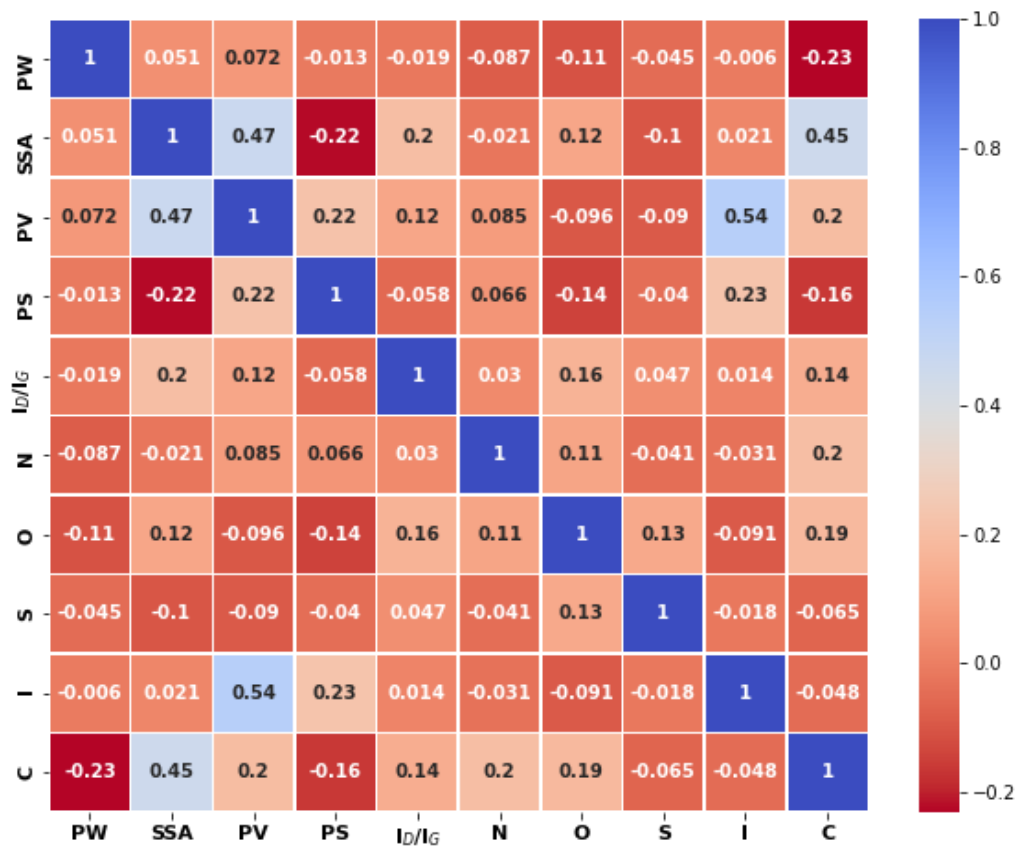

Figure 3: Spearman's correlation between the physicochemical characteristics of the carbon electrodes of supercapacitors in the considered dataset.

## 3.2. Comparison between regression methods

After completing data profiling and correlation analysis, we applied five different regression models to the dataset: ordinary least square regression, support vector regression, decision tree, random forest, and extreme gradient boosting. The dataset was divided into two parts: 70% of the data were randomly selected for training the regression models and the remaining 30% for testing. The total number of data entries used for training and testing was 3176 and 1362, respectively. The nine physicochemical characteristics of the carbon electrodes of supercapacitors in the dataset were considered as independent variables, while the resulting specific capacitance as the dependent one. The results are depicted in Figure 4 as a comparison between the real specific capacitance ($C_{Real}$) obtained from the literature articles in the dataset and the values of specific capacitance predicted by the regression models ($C_{ML}$). In each panel, the perfect match between the actual specific capacitance and the predicted one is shown via the straight diagonal line, where $C_{Real} = C_{ML}$.

As illustrated in Figure 4 (a), the matching between the actual and predicted values of specific capacitance using OLS method was low, as evidenced by the significant deviation of numerous data points from the diagonal line and the $R^2$ value of 0.32. Additionally, the large *RMSE* and *MAPE* values in Table 2 indicate that OLS regression achieves inferior prediction capability when compared with DT, SVR, RF, and XGBoost approaches.

The performance analysis of the tree-based model indicates that the DT model in Figure 4 (c) does not accurately predict the actual specific capacitance. Instead, the SVR, RF and XGBoost models were more accurate at predicting the specific capacitance. As illustrated in Figure 4 (b), (d) and (e), most data points lie near the diagonal line, indicating prediction accuracy as supported by the $R^2$ values of 0.72, 0.75 and 0.79 for the SVR, RF and XGBoost models, respectively. Moreover, other performance parameters (*RMSE*, *b'* and *MAPE*) also indicate that the SVR, RF and XGBoost models yielded superior regression capabilities when compared to the OLS and DT models. Since the performance analysis in Table 2 revealed that the XGBoost model showed the best $R^2$ and *RMSE* values, only this regression was employed in the following analyses on the dataset. Notice that XGBoost showed better prediction performance than an artificial neural network as well (see Supplementary Note 4 for details).

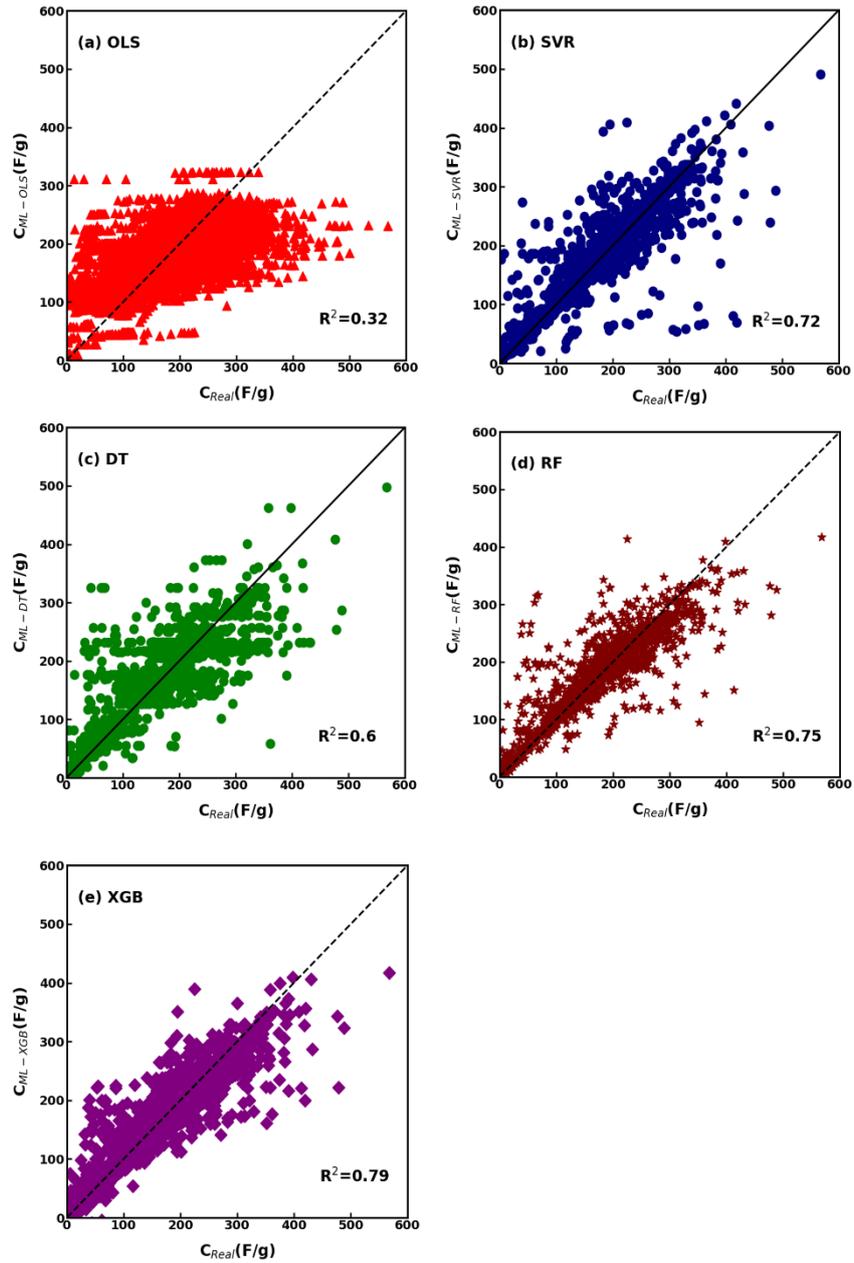

Figure 4: Comparison between the actual specific capacitance from literature research articles in the dataset and the predicted specific capacitance from (a) OLS, (b) SVR, (c) DT, (d) RF, and (e) XGBoost models.

| Model | $R^2$ | RMSE | b' | MAPE |
|---|---|---|---|---|
| OLS | 0.32 | 71.52 | 1.00 | 100.62 |
| SVR | 0.72 | 46.3 | 0.98 | 31.59 |
| DT | 0.60 | 55.63 | 1.00 | 36.28 |
| RF | 0.75 | 43.96 | 0.98 | 27.09 |
| XGBoost | 0.79 | 40.27 | 0.95 | 30.08 |

Table 2. Performance analysis of the different regression models on the collected dataset of carbon-based supercapacitors.

## 3.3. Influence of specific capacitance testing method

It is well established that the method of experimental testing can influence the magnitude of specific capacitance of supercapacitors. For instance, for the AC-based electrode developed by Meng *et al.*,[68] specific capacitance values of 225 F/g and 465 F/g were measured using two-electrode and three-electrode testing methods, respectively. Therefore, to investigate the effect of testing method on specific capacitance, the primary dataset generated in the current study was divided into two different subsets, wherein one contained the specific capacitance values obtained using the three-electrode method of testing (2754 data entries) and the other comprised the specific capacitance values obtained using the two-electrode method (1784 data entries). The XGBoost model was trained again on each of these two subsets of data.

The actual specific capacitance and the predicted results by XGBoost model were found to strongly match for the three-electrode method of testing, as evident in Figure 5 (a). Such good prediction capability is also highlighted by the statistical performance parameters *viz.* $R^2 = 0.89$, $RMSE = 28.71$, $b' = 0.98$, and $MAPE=28.71$, being improved with respect to the regression analyses on the whole dataset reported in Section 3.2. Hence, the testing method has a significant effect on the specific capacitance value, being a further (categorical) variable to be considered in the prediction of specific capacitance. Thus, we investigated the significance of the different independent variable on the trained XGBoost model, which indicates how the physicochemical parameters of carbon electrodes influence the specific capacitance. Figure 5 (b) depicts the feature importance analysis for the three-electrode testing method, where higher shares are associated to more influence of variables on specific capacitance: the SSA, heteroatom doping (N%), and PV were found to be the major factors influencing the specific capacitance.

The correlation between the actual and predicted specific capacitance for the datasets obtained using the two-electrode method of testing is shown in Figure 5 (c), instead. Again, the regression accuracy is improved by considering the subset of data measured with two-electrode method rather than the whole dataset. In fact, the performance parameters for the XGBoost regression of the two-electrode dataset are $R^2 = 0.93$, $RMSE = 19.45$, $b' = 0.989$, and $MAPE = 31.07$, thus better with respect to the analyses carried out on the whole dataset. In this case, the PW, SSA, and the $I_D/I_G$ ratio of the carbon electrode were found to contribute most towards enhancing the supercapacitors' specific capacitance, as observed from the feature analysis in Figure 5 (d). Interestingly, the SSA is found as an influential physicochemical characteristic of supercapacitors in both testing methods, while small discrepancies emerge for the other variables. PW is found as a relevant parameter of specific capacitance only in case of two-electrode measures, thus appearing as a possible descriptor able to discriminate between the adopted method of testing.

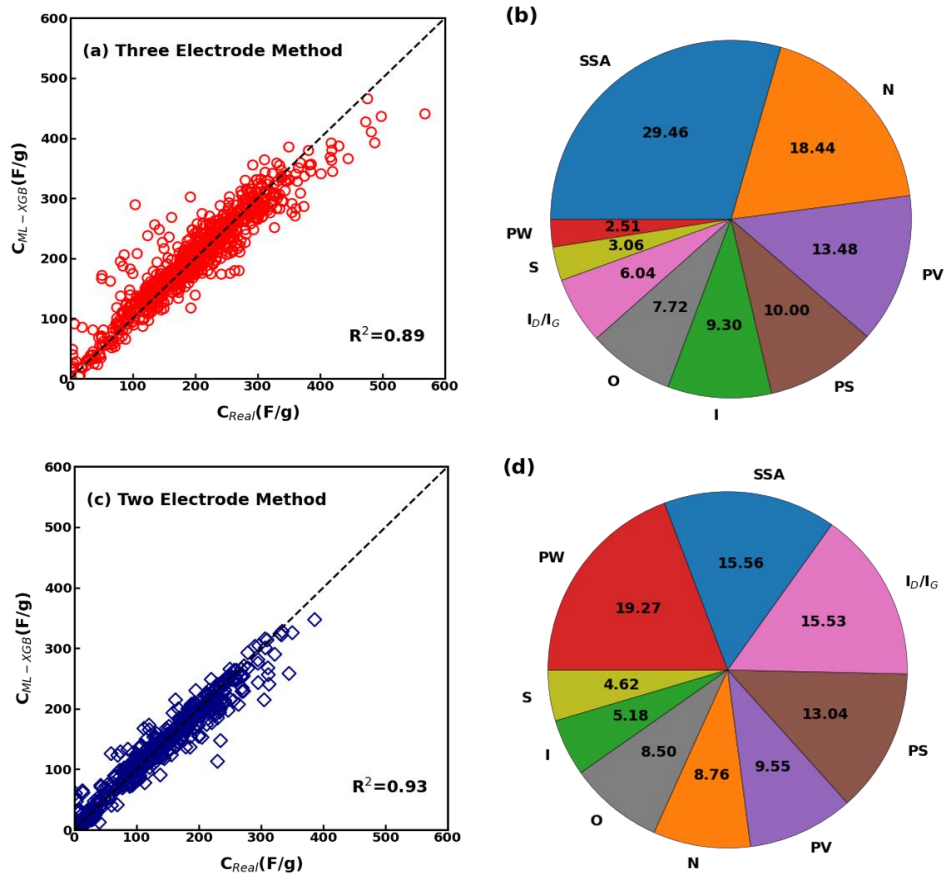

Figure 5: (a) Comparison between the predicted and actual capacitance values obtained by the three-electrode method; (b) feature analysis for the three-electrode method of testing. (c) Comparison between the predicted and actual values for the two-electrode testing method; (d) feature analysis for the two-electrode method of testing.

### 3.4. Influence of electrolyte

The specific capacitance of a supercapacitor is determined not only by the physicochemical behaviour of the electrode material and the testing method, but also by the type of electrolyte used. For instance, Zhou *et al.* [52] synthesised hierarchical nitrogen-doped porous carbon and demonstrated a specific capacitance of 339 F/g in 6M KOH and 282 F/g in 1 M $H_2SO_4$, respectively, at a current density of 0.5 A/g. Thus, to decouple the effect of different electrolytes from our analyses, we considered configurations with either 6M KOH or 1M $H_2SO_4$. Consequently, we extracted only data entries characterized by 6M KOH (2819 entries) and 1M $H_2SO_4$ (471 entries) from the overall dataset. In these cases, 80% of dataset was considered for training and 20% for testing the XGBoost model. The accuracy of the obtained regression is corroborated by the improved statistics of XGBoost model fitting for the 6M KOH electrolyte ($R^2$ = 0.81, RMSE = 33.86, b' = 1.01, and MAPE = 16.75) and the 1M $H_2SO_4$ electrolyte ($R^2$ = 0.87, RMSE = 36.48.44, b' =0.98, and MAPE =116.24), as depicted in Figure 6 (a) and 6 (c). The feature analyses carried out on supercapacitors with 6M KOH and 1M $H_2SO_4$ electrolytes demonstrate that the SSA, nitrogen doping, and PV were the major contributors to the capacitive performance in the 6M KOH electrolyte, whereas the SSA, nitrogen doping, and PW in the 1M $H_2SO_4$ electrolyte, as evident from Figure 6 (b) and 6 (d).

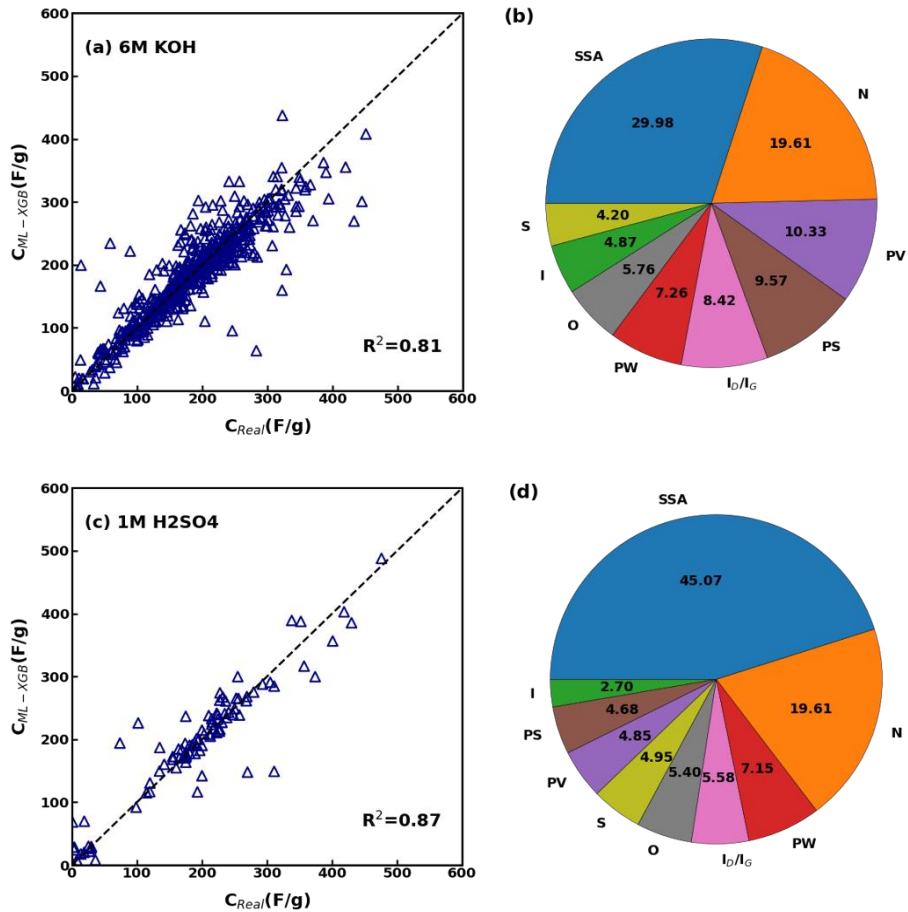

Figure 6: (a) Comparison between predicted and actual specific capacitance and (b) feature analysis for the subset of data having 6M KOH electrolyte. (c) Comparison between predicted and actual specific capacitance and (d) feature analysis for the subset of data having 1M H$_2$SO$_4$ electrolyte.

Due to the limited data entries for the 1M H$_2$SO$_4$ electrolyte, then we considered only the aqueous electrolyte 6M KOH – which has also the additional benefits of being inexpensive, safe, and with a high dielectric constant and specific capacitance – to discriminating again between two-electrode and three-electrode testing methods. Refining the datasets considering a specific electrolyte (6M KOH) and method of testing further improved the regression performance with respect to results in Sections 3.2 (overall dataset) and 3.3 (datasets separated for three- and two-electrode testing methods). This is evident from Figure 7 (a) and (b), where most data points are located near the diagonal line, indicating a strong correlation between the actual and predicted specific capacitance values. Furthermore, the accuracy of regression is corroborated by the improved statistics of XGBoost model fitting for both the two- ($R^2$ = 0.95, *RMSE* = 16.56, *b'* = 0.97, and *MAPE* = 10.15) and the three-electrode method ($R^2$ = 0.91, *RMSE* = 24.08, *b'* =0.985, and *MAPE* =12.88). Similarly to Figure 5, the feature analysis carried out on supercapacitors with 6M KOH electrolyte demonstrate that the SSA, PV, and nitrogen doping were the major contributors to the capacitive performance in the three-electrode testing method (Figure 7 (b)), whereas the PW, PS, and SSA in the two-electrode one (Figure 7 (d)). Hence, the regression performed on the limited dataset of supercapacitors with 6M KOH electrolyte does not change the relative influence of physicochemical characteristics of supercapacitors on their performance discussed in Section 3.3, therefore showing the robustness of the feature analysis.

Notice that, due to limited data entries for different concentrations of electrolyte in the current database (1M KOH has 12 entries, 2M KOH has 169 entries and 3M KOH has 132 entries), we could not train a robust XGBoost model specifically dedicated to exploring also this effect on the specific capacitance of supercapacitors.

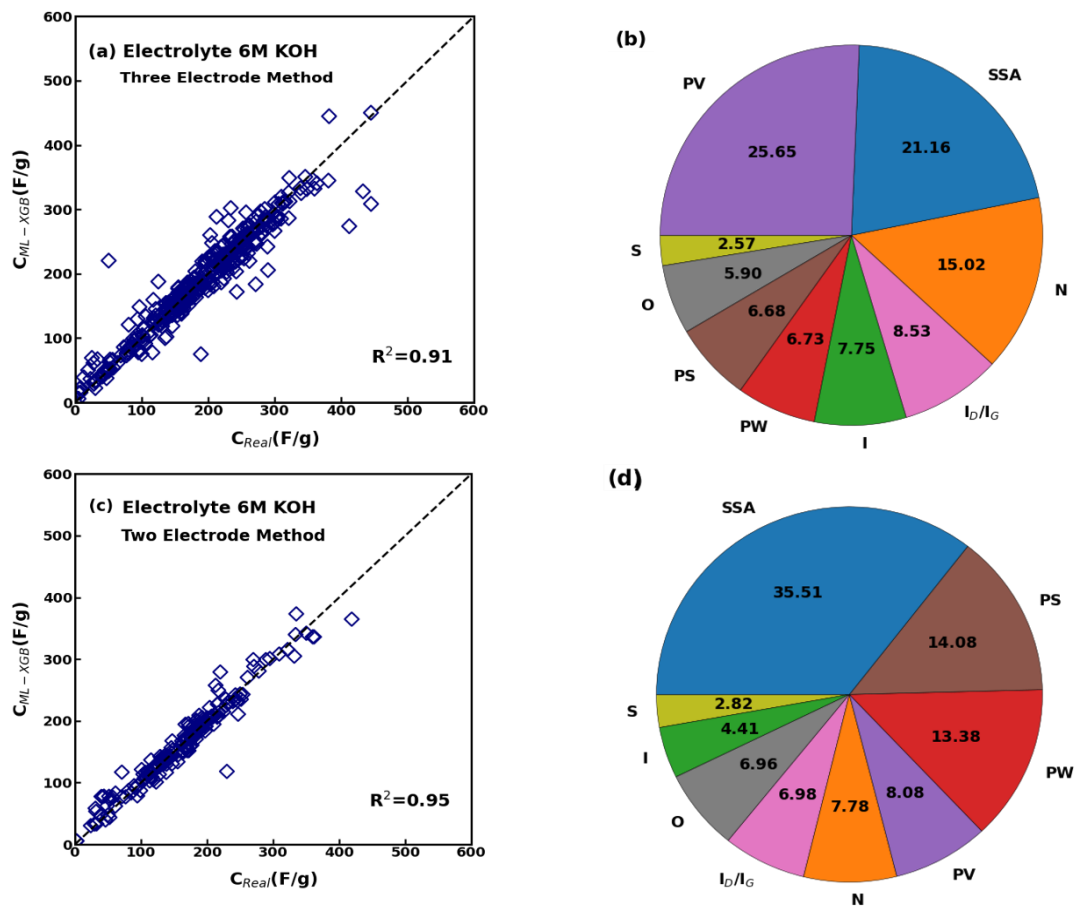

Figure 7: (a) Comparison between predicted and actual specific capacitance and (b) feature analysis for the subset of data having 6M KOH electrolyte and measured by three-electrode method. (c) Comparison between predicted and actual specific capacitance and (d) feature analysis for the subset of data having 6M KOH electrolyte and measured by two-electrode method.

### 3.5. Influence of carbon electrode structure

Carbon exists in various allotropic forms with distinct morphologies and physicochemical properties. Activated carbon, hierarchical porous carbon, heteroatom doped porous carbon and graphene derived carbon are mainly employed for carbon electrodes of supercapacitors. The different morphological forms of these carbon allotropes may affect the specific capacitance of supercapacitors. [28] ACs possess a large SSA and pore volume, thus easing the accumulation of static charges at the electrode surface and the resulting specific capacitance. HPC electrodes, instead, contain pores in a wide range of length scales (from micro to macro). The presence of macropores in HPC allows high-rate ion transport and acts as an ion reservoir. HA carbon electrodes are generally obtained from AC incorporated with heteroatoms (N, O, S, P), which enhance the wettability and electronic conductivity of the base material. Graphene shows also high SSA and electrical conductivity.

Therefore, we further split our dataset according to the type of carbon materials used in the construction of the electrodes. AC, HPC, and heteroatom (HA)-doped electrodes were differentiated to generate separate datasets, and XGBoost trained to best match the capacitive behaviour of supercapacitors made of specific carbon structures. Notice that, in this case, datasets have not been subdivided according to different testing methods or electrolytes, since the limited number of entries available for some classes of carbon structures did not allow a robust training of the regressor. Figures 8 (a), (c) and (e) compare the predicted and actual values of specific capacitance for the three types of considered carbon structures, highlighting a good match especially for HA ones (correlation statistics are detailed in Table 3). The feature analysis is done also in this case: Figures 8 (b), (d) and (f) identify as most influential physicochemical features a combination of parameters previously found for two- and three-electrode testing methods, such as PW, SSA, and N%.

| Electrode type | $R^2$ | RMSE | $b'$ | MAPE |
|---|---|---|---|---|
| AC | 0.82 | 33.85 | 1.003 | 23.104 |
| HPC | 0.77 | 44.13 | 0.97 | 49.543 |
| HA | 0.9 | 26.78 | 0.98 | 16.276 |

Table 3. Performance analysis of the trained XGBoost models to relate the physicochemical features of carbon electrodes with different structures and the measured specific capacitance.

Overall, the current study focused mainly on the results obtained by applying XGBoost regressors, as their accuracy was shown to be superior to that of other ML models. As a result of differentiating datasets according to the testing method, electrolyte type or morphology of the carbon electrode material, the accuracy of trained XGBoost models was further improved (see $R^2$ values), while the most relevant physicochemical features identified for these different categorical variables. Considering all the feature analyses shown in Figures 4, 5, 6, 7, and 8, SSA is by far the dominant physicochemical characteristic of electrodes in determining the specific capacitance of the supercapacitors in the dataset, followed by N% and PW (which appears to be particularly influent when two-electrode methods are employed for the measure). PV, $I_D/I_G$ and PS follow with decreasing importance, while I, O% and S% result to be the least influential features.

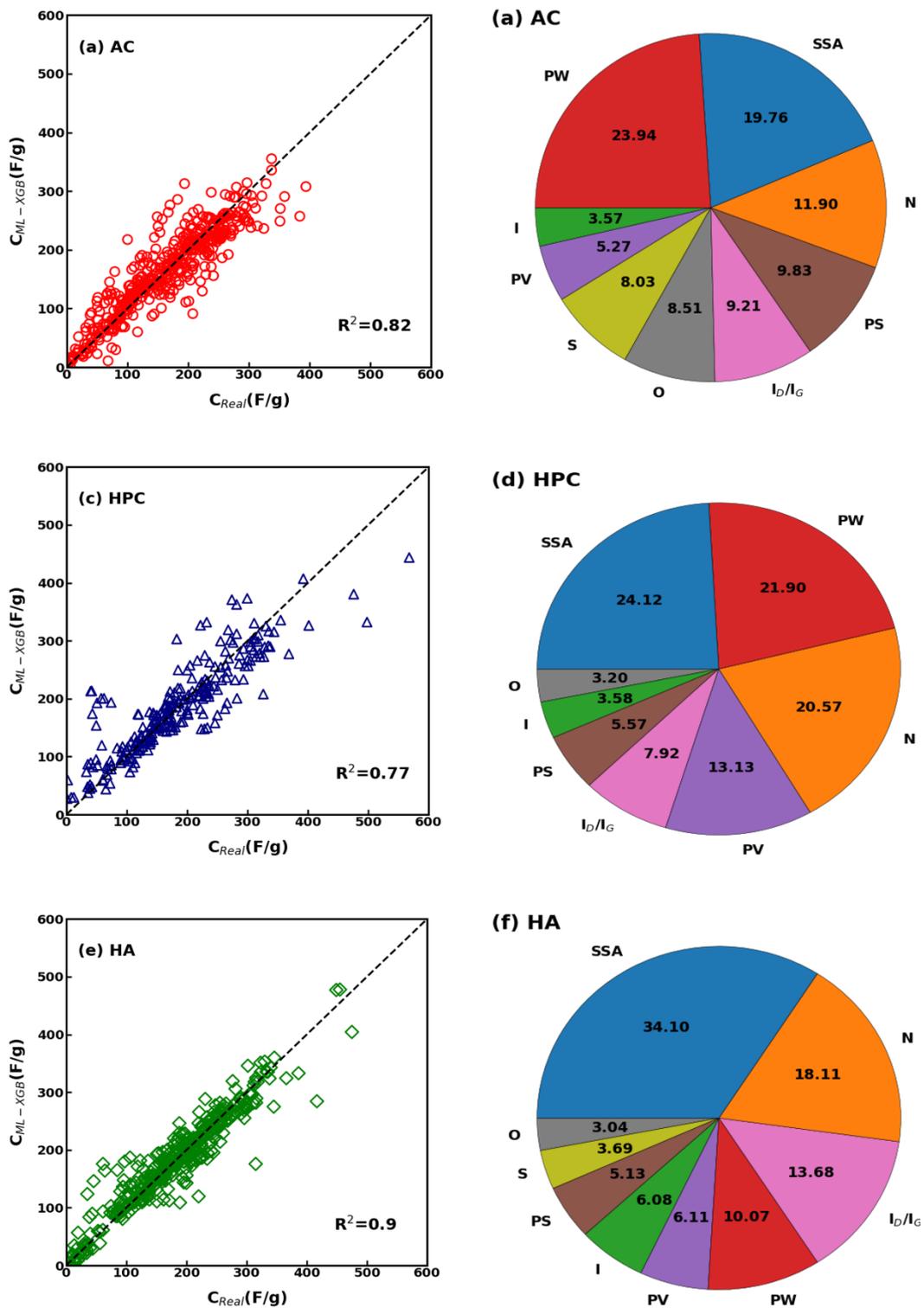

Figure 8. (a) Comparison between the predicted and actual specific capacitance and (b) feature analysis for the activated carbon electrodes. (c) Comparison between the predicted and actual specific capacitance and (d) feature analysis for the hierarchal porous carbon electrodes. (e) Comparison between the predicted and actual specific capacitance and (f) feature analysis for the heteroatom-doped carbon electrodes.

## 4. Conclusions

ML models such as OLS regression, SVR, DT, RF, and XGBoost were used to predict the influence of various physicochemical parameters and categorical variables of carbon-based electrode materials on the capacitive performance of an ELDC supercapacitor. First, a dataset was developed by extracting information from 147 experimental research articles on carbon-based electrode supercapacitors. This included the presence of defects, the pore volume, pore surface, current density, surface specific area, potential window, nitrogen, oxygen, and sulphur content in carbon-based electrode materials. Categorical variables such as the testing method, electrolyte type or morphology of the carbon electrode material were also considered. These data entries (4538) were fed into five regression models, prior to which the dataset was curated to achieve consistent physical units and outlier detection. Subsequently, the Spearman's rank correlation coefficient was used to determine the correlation between each pair of parameters, which suggested that all the available physicochemical parameters were not dependent from each other. For training the regression models, the datasets were divided into a 70:30 ratio for training and testing, respectively. Correlations between the actual specific capacitance and the predicted specific capacitance of the five models are ranked as follows: XGBoost>RF>SVR>DT>OLS, thus showing a superior regression performance by ML algorithms.

Additionally, we used the XGBoost model to predict the effect of the testing method (two- and three-electrode method) on the specific capacitance of supercapacitors. This resulted in acceptable performance parameters for both the testing methods. Furthermore, in the three-electrode method, SSA, N%, and PV were identified as the major contributors, whereas in the two-electrode method SSA, PW, and $I_D/I_G$ were observed to significantly influence the capacitive performance. To comprehend the impact of the electrolyte on the specific capacitance, we further extracted datasets having 6M KOH in the two-electrode and three-electrode testing methods. The performance parameters obtained using the XGBoost method suggested improved statistics for both the testing methods. As a result, the PV, SSA, and N% were identified as the significant contributors in the three-electrode method, whereas SSA, PS, and PW were confirmed to be the significant contributors in the two-electrode method. Finally, using the XGBoost model, we determined the various physicochemical characteristics according to the type of electrode materials used for the construction of the electrode that affect the specific capacitance of supercapacitor. The heteroatom (HA)-doped carbon exhibited a better regression in comparison to the AC and HPC. Overall, SSA appears as the most influential physicochemical characteristic of electrodes in determining the specific capacitance, followed by N% and PW. PV, $I_D/I_G$ and PS have a decreasing importance, while I, O% and S% the least.

We highlight that the imperfect matching between the currently trained ML models and the considered experiments may be also due to different experimental conditions during the supercapacitor testing. For instance, electrode conditioning before property measurements usually involves extensive charge-discharge cycling or holding the electrode for some time at an elevated temperature and potential, and it typically decreases or even removes certain surface functional groups, which in turn improves storage stability. Unfortunately, not all researchers perform electrode conditioning before measuring properties or even identify possible current leakages when reporting measurements. This study, however, is not intended to replace modelling or experimental analyses, but rather to provide a preliminary support in the design and development of new supercapacitors, with the possibility to re-train and thus refine the presented ML models as soon as further data and descriptors will become available in the literature (e.g., extensive data on carbon material percentage in the electrode, cf. Supplementary Note 5).

In conclusion, the current study demonstrated the successful utilization of a data-driven method to predict the material performance for supercapacitor applications and revealed the most significant parameters that affect the specific capacitance of EDLCs. In perspective, the curated dataset developed and shared in this work may facilitate further analyses and potential optimization of carbon-based electrodes in different electrochemical applications. To ease the exploitation of the trained models (see Supplementary Note 6) by experimentalists, we developed a software with a graphical user interface (SUPERCAPs [69]) that allows to easily provide with an estimate of the specific capacitance of carbon-based supercapacitors knowing the physicochemical characteristics and structure of carbon electrodes, the testing method, and electrolyte.

**Acknowledgements:** We thank Dr. Paolo Bondavalli for helpful discussions.
**Author Contributions:** Conceptualization and methodology by R.S.; data collection and curation by A.A; software development and testing by S.M.; results visualization and original draft preparation by A.M.; supervision, editing and formal analysis performed by M.F., E.C. and P.A. All authors reviewed the manuscript.
**Funding**: This work has received funding from the European Unions' Horizon2020 research and innovation programme "SMARTFAN: Smart by Design and intelligent by Architecture for Turbine Blade Fan and Structural Components System" [grant agreement number 760779].
**Data availability statement**: All data collected and analyzed during this study are included in this published article, the related supplementary information files and Zenodo archive (10.5281/zenodo.7346943).
**Competing interests:** The authors declare no competing interests.
**Correspondence** and request for materials should be addresses to M.F.
**Published article**: The editorial published version of this article is available at https://doi.org/10.1038/s41598-023-33524-1

# The impact of physicochemical features of carbon electrodes on the capacitive performance of supercapacitors: A machine learning approach

# - Supplementary Information -

Sachit Mishra[a,b,1], Rajat Srivastava[a,c,1], Atta Muhammad[a,d], Amit Amit[a], Eliodoro Chiavazzo[a], Matteo Fasano[a,*], and Pietro Asinari[a,e]

[a] Department of Energy "Galileo Ferraris", Politecnico di Torino, Corso Duca degli Abruzzi 24, 10129 Torino, Italy.
[b] IMDEA Network Institute, Universidad Carlos III de Madrid, Avda del Mar Mediterraneo 22, 28918 Madrid, Spain.
[c] Department of Engineering for Innovation, University of Salento, Piazza Tancredi 7, 73100 Lecce, Italy.
[d] Department of Mechanical Engineering, Mehran University of Engineering and Technology, SZAB Campus, Khairpur Mir's 66020, Sindh, Pakistan.
[e] Istituto Nazionale di Ricerca Metrologica, Strada delle Cacce 91, 10135 Torino, Italy.
* Corresponding author.
Email address: matteo.fasano@polito.it (Matteo Fasano).
[1] *Equal contributors.*

**Supplementary Note 1: Review on carbon electrodes**

*Porous and activated porous carbon*

Because of their high specific surface area (SSA), improved electrical conductivity, adjustable pore-sizes, electrochemical stability, and low cost, activated porous carbons (AC) offer significant promise for use as the electrode [1-4]. These properties make AC an excellent material for a variety of applications, including water purification, gas separation and storage, and electrode materials for capacitors, fuel cells and batteries. Porous/activated carbons are prepared by the pyrolysis of petroleum coke and coal followed by physical/chemical activation [5]. In recent years, synthesising AC from fossil raw materials has been highly discouraged and several efforts were made to find sustainable and green sources of raw materials for the AC preparation. In this direction, biological matters such as roots, flowers, stems, leaves, fungi fruits, and animal body parts etc., have been adopted as resources for the synthesis of AC [6]. Additionally, several methodologies have been used to synthesise the AC from the biomass precursor including pyrolysis, hydrothermal carbonization, mechano-chemical, hard- and soft-templating [7].

In pyrolysis, the biomass is heated at an elevated temperature (T = 300–1000 °C) under the presence of inert gas (e.g., nitrogen, argon). The conversion of biomass into AC involves several steps including removal of moisture (T<100 °C), degradation of cellulose and hemicellulose (in temperature range 200 °C<T<500 °C), and the decomposition of lignin (T>500 °C). The carbonization of biomass removes all the volatile materials, and a major part of the residual solid is carbon [8]. Additionally, the application of pyrolysis-derived carbons (DPCs) in energy storage devices depends on their pore morphology and physicochemical characteristics.

Usually, DPCs have a low SSA and unsuitable pore morphology for their application in supercapacitors. Thus, an activation process (physical or chemical activation) is needed to enhance the physicochemical characteristics of DPC. In the case of physical activation, the carbon precursor is carbonized at a temperature (T<800 °C) followed by an activation process in the existence of activating agents such as air, $CO_2$, and steam at an elevated temperature. The chemical activation process requires the mixing of carbon precursor with an activating agent at a suitable temperature [9]. Because of its lower activation temperature, high yield, and generation of microporous carbon with a large SSA, potassium hydroxide (KOH) is used as one of the most common chemical activating agents.

In hydrothermal carbonization (HTC), the carbonization process of biomass occurs at natural conditions, i.e., in presence of an aqueous medium at mild temperature (130–250 °C) and pressure (0.1 MPa). This process is very complex and usually contains five steps: hydrolysis, dehydration, decarboxylation, polymerization, and aromatization [10]. Additionally, in HTC processes, the nature and morphology of the AC can be tuned by varying the temperature under mild processing conditions [10].

In general, ACs possess a large SSA (> 1000 $m^2$/g) and pore-volume (> 0.5 $cm^3$/g), which are critical characteristics because the accumulation of static charges at the electrode surface determines the charge storage capacity of the supercapacitor. However, in practical observation, the capacitance of the ACs electrode supercapacitor is only about 10–20% of its theoretical capacitance [11]. This is either due to the presence of inaccessible micropores or very large pores generated during carbon activation. Therefore, the high capacitive performance of the supercapacitor depends on the characteristics of electrode materials such as large SSA and optimal pore-size distributions to ease the transport of electrolyte ions within the pores [11].

The cost of AC hinders its application in a supercapacitor as an electrode material. Recent studies reported that the AC can be successfully synthesised using various biological sources and wastes such as pitaya peel [12], corncob residue [13], water bamboo [14], a harmful aquatic plant (*Altemanthera philoxeroides*) [15], pumpkin [16], pomelo peel [17]. They are cost-effective, environmentally friendly, in abundance, and renewable [18]. One such low-cost bio-waste is rice husk produced during the processing of rice. The annual production of rice husk is about 120 million tonnes all over the world [19] and its disposal is a serious environmental concern as the most common disposal method is the open-air burning of rice husk in an uncontrolled environment, which releases a large amount of carbon monoxide (CO) and carbon dioxide ($CO_2$) [20]. Cellulose, hemicellulose, silica, lignin, and moisture are the main constituent of the rice husk [21].

Guo *et al.* [22] used rice husk to synthesise AC with high SSA (ranging from 1392–2721 $m^2/g$) by using alkaline hydroxide (KOH and NaOH) activation. They reported that the EDLC with rice husk AC electrode could achieve the specific capacitance of 210 F/g in the KCL solution. Furthermore, a high-performance supercapacitor using rice husk derived AC electrode developed by He *et al.* [23] obtained a capacitance of 245 F/g in a current density of 0.05 A/g, along with a slight decrement in the charge storage capacity (233 F/g) under increasing current densities (2 A/g). Wang *et al.* [24] also synthesised a AC electrode from hydrochar derived rice husk by KOH activation, which yielded a high specific capacitance of 312 F/g due to the high SSA (3362 $m^2/g$) of the activated rice husk hydrochar electrode. Additionally, Chen *et al.* [25] synthesised a AC electrode for EDLC supercapacitor with a SSA of approximately 4000 $m^2/g$, which obtained a specific capacitance of 368 F/g in a 6 M KOH electrolyte.

Similar to rice husk, corn stalk core is also a bio-waste that shows the possibility for the generation of AC electrodes for supercapacitors. The corn stalk core contains natural pores distributed in a sponge-like structure that are suitable for preparing the high surface area and AC electrode raw material. Yu *et al.* [26] prepared ACs electrode using corn stalk core with a high surface area (2349.89 $m^2/g$) and determined the specific capacitance of 140 F/g (at 1 A/g current density). Additionally, Wang *et al.* [27] reported the conversion of corncobs to activated carbon using chemical activation for application in a supercapacitor. The resultant ACs exhibited a high SSA of 3054 $m^2/g$, and specific capacitances of 401.6 F/g and 328.4 F/g in 0.5 M $H_2SO_4$ and 6 M KOH electrolyte, respectively, at a current density 0.5 A/g. Karnan *et al.* [28] fabricated a supercapacitor device using an activated carbon electrode derived from corncobs and an ionic liquid electrolyte. With just a ten-second charge, the device could power a LED light for more than 4 minutes. The electrochemical performance of activated carbon (AC) electrodes in 0.5 M $H_2SO_4$ electrolyte also revealed the high capacity of the corncob electrode with a specific capacitance of 390 F/g at 0.5 A/g [28]. Several authors also developed activated carbon (AC) electrodes from corn-based biomass such as corn straw, corn gluten, popcorn, etc. with improved SSA as well as reasonable electrochemical performance [29-31].

In a recent study, Rajabathar *et al.* [32] prepared a porous AC nanostructured electrode using jackfruit peel waste (JFPW), which showed outstanding electrochemical performance with a specific capacitance of 320 F/g at low current density (1 A/g) in 1 M $Na_2SO_4$ electrolyte. They also reported that AC electrodes derived from jackfruit retained a high specific capacitance 274 F/g even at a high current density (5 A/g). Similar evidence of a high-performance supercapacitor was also reported using AC electrode prepared from pitaya peel having a specific capacitance of 255 F/g at a current density 1 A/g and 96.4% retention capacity at 5 A/g with excellent stability in 6 M KOH electrolyte [12]. Additionally, Lin *et al.* [17] synthesised hemicellulose AC raw material from pomelo peel for an electrode with a high SSA of 1361 $m^2$/g and excellent charge storage capacity of 302.4 F/g at a current density of 0.5 A/g.

Three-dimensional sakura-based activated carbons have also been utilized as the raw material for the electrode of supercapacitor. The electrochemical performance was analysed in a three-electrode method of testing with 6 M KOH electrolyte, reporting a specific capacitance of 265.8 F/g for sakura-based active carbon at a current density 0.2 A/g [33]. Chang *et al.* also synthesised activated ACs using paulownia flower as the precursor, which offered 297.1 F/g specific capacitance at 1 A/g in 1 M $H_2SO_4$ electrolyte [34]. Zhang *et al.* [35] reported the conversion of bamboo through carbonization and chemical activation into activated porous carbon for supercapacitor electrodes with a high specific capacitance of 293 F/g at 0.5 A/g current density. Wang *et al.* [36] modified commercially available coconut shell-based AC (CSAC) through $H_2O$ plasma resulting in an environmentally friendly method to generate electrode material (HCSAC) with enhanced specific capacitance and retention capability compared to its precursor CSAC.

Activated carbon-based supercapacitor electrodes have also been synthesised using a low cost, highly porous willow-wood. The resultant ACs exhibit a high surface area (2793 $m^2$/g), pore-volume (1.45 $cm^3$/g) and the presence of both micro- and meso- pores that are favourable for the energy storage [37]. The obtained AC electrode supercapacitor also showed a magnificent electrochemical performance having a specific capacitance of 394 F/g at a current density of 1.0 A/g in an aqueous electrolyte (6 M KOH) [37].

One socio-economic and environmental concern worldwide is to get rid of seaweed (*Ascophyllum nodosum*) blooms, which is in abundance in northern oceans. Chemically activated biocarbon electrode derived from *Ascophyllum nodosum* can also be used in supercapacitors. Perez-Salcedo *et al.* [38] synthesised an activated biocarbon electrode derived from *Ascophyllum nodosum* for supercapacitors with a capacitance of 207.3 F/g at current density (0.5 A/g), excellent stability and retention capability.

*Hierarchical porous carbon*

The hierarchical porous carbon (HPC) material contains pores in a wide range of length scales that are missing in the conventional porous material. The HPC contains pores in the ranges of macro- (>50 nm), meso- (2–50 nm), and micro- (<2 nm) scales. The presence of macropores in HPC allows high-rate ion transport and acts as an ion reservoir. Furthermore, the interconnected mesopores provide low resistance pathways for the diffusion of ions, and the high SSA of micropores, which enhances the adsorption of ions at the pore surface [39]. These unique properties of HPC material gained recent interest in the selection of electrode material for supercapacitors. The development of hierarchical porous structure from carbon material requires templating techniques. There are two types of templating: soft template; hard template. The soft templates were used as a substance for self-assembly; whereas the hard template method consists of the three steps: 1) impregnating the pre-synthesised template; 2) carbonization; 3) peeling off the hard template. It is followed by the carbon conversion process (carbonization) and etching [39].

However, these methods are complex, time demanding, and expensive. In addition, it is hard to regulate porosity that creates a serious obstruction in the usage and large-scale production of HPC. Thus, there is a special need for the development of an easy, inexpensive, and eco-friendly procedure for the synthesis of HPC materials.

The synthesis of AC material from natural biomass such as cotton is an eco-friendly and economical approach for the development of electrodes for supercapacitors. The bio swelling of cotton fibres under the influence of NaOH/urea enables the formation of HPC fibres with improved surface characteristics [40]. The cotton fibre derived HPC fibres electrode material possess a high SSA (584.49 $m^2$/g), along with favourable pore morphologies that enhance the specific capacitance (221.7 F/g at 0.3 A/g) [40].

Bagasse (a biomass-waste from sugar industries) were used to synthesise carbonaceous material for the absorption of heavy metal ions, organic pollutants, and it finds its applicability in energy storage application. However, the carbonaceous material derived from bagasse contains narrow pore-size distribution and insufficient SSA that restricts its applicability in supercapacitors. Feng et al. [41] developed a simple method for the synthesis of HPC from bagasse using sewage sludge assisted hydrothermal carbonization with KOH activation. This process is a cheap and efficient way to regulate the porosity and structure of HPC and results in an excellent supercapacitive performance. The bagasse-derived hierarchical structured carbon (BDHSC) electrode supercapacitor possesses 320 F/g capacity at 0.5 A/g current density with good cyclic stability. Zhou et al. [42] synthesised the nitrogen-doped porous carbons (HNPCs) from biomass precursor cellulose carbamate with tuneable pore structures and ultrahigh SSAs via simultaneous carbonization and activation using a facile one-pot approach. They exhibited ultrahigh specific surface area (3700 $m^2$/g), high pore-volume (3.60 $cm^3$/g) and high-level nitrogen doping (7.7%). In three-electrode system, HNPCs showed specific capacitance of 339 F/g with 6 M KOH electrolyte and 282 F/g with 1 M $H_2SO_4$ electrolyte at a current density of 0.5 A/g, whereas in two-electrode system it exhibited a high specific capacitance of 289 F/g at 0.5 A/g.

Gou et al. [43] prepared a HPC material for the electrode of the supercapacitor from wheat straw cellulosic foam with high SSA of 772 $m^2$/g after KOH activation and micropores ranging from 1.05-1.74 nm with 6 M KOH electrolyte. The high porosity provides the better migration of the ions in an electrolyte, thus enhancing the electrochemical characteristics of the capacitors. In three-electrode system, they obtained a capacitance of 226.2 F/g at a current density of 0.5 A/g. This provides a method for obtaining electrode materials from the cheap and easily available material wheat straws.

In a recent study, Zhao et al. [44] prepared N-O co-doped AC from low-cost, sodium alginate particles for the development of supercapacitors. They obtain the N-O co-doped AC from the crosslinking of SA beads with diammonium chains with the help of electrostatic interaction between ammonium cations and carboxylate groups of SA chains. Both the species (N-O) and concentration of diammonium chains strongly affected the electrochemical characteristics of NO co-doped AC. They obtained capacitance performance of 269.0 F/g at current density of 1 A/g.

*Heteroatom doped porous carbon*

Besides the SSA and pore-volume, there are also several other factors that influence performance such as surface functional groups and conductivity (pseudocapacitance and overall electric capacity). These can be achieved by the appropriating carbon with heteroatoms (nitrogen, oxygen, sulphur, etc.). By doing so, it not only enhances the wettability but also improves electronic conductivity of AC [45]. Among these heteroatoms (N, O, S, P), sulphur is the highest reactive element. Sulphur doping of carbonaceous material increases its band-gap that enhances the electron donor properties and change in the electronic density of state. Sulphur doping also increases wettability, which in turn decreases the diffusion resistance that occurs between the electrode and electrolyte ions [46].

Li *et al.* [47] used willow catkin to develop porous carbon nanosheets (PCNs) from pyrolytic and activation approaches, followed by the co-doping of nitrogen and sulphur. The N-S co-doped carbon nanosheets electrode supercapacitor possesses 298 F/g capacity at 0.5 A/g current density and 298 F/g capacity at 0.5 A/g with green and low-cost materials for electrode of supercapacitor. Wang *et al.* [45] reported a porous nitrogen self-doped carbon material with layer structure for high-performance supercapacitors. It was derived from the by-product of the pig-farming industry porcine bladders. It possesses C, N and O elements in abundance. Combining carbonization and KOH activation processes yielded the nitrogen self-doped layered AC. KOH dosage can be changed to adjust the amount of N and pore structure. It has outstanding electrochemical characteristics including high specific capacitance of 322.5 F/g and good cycle stability during 5000 cycles.

Kim *et al.* [43] reported a straightforward method for biomass-derived AC with a high surface area and heteroatom doping. It involves exothermic pyrolysis of Mg/K/MgK-nitrate-urea-cellulose mixture followed by a high temperature carbonization and washing treatment. The produced N-doped AC material shows specific capacitance of 279 F/g at 1 A/g in 6 M KOH electrolyte in the two-electrode method of testing. Wan *et al.* [48] prepared three AC from lotus pollen for supercapacitor activated with ZnCl2, $FeCl_3$, and $CuCl_2$. AC obtained by $CuCl_2$ activation exhibits higher surface area, more porous and higher heteroatom doped than traditional activated $ZnCl_2$ or $FeCl_3$ AC.

Demir *et al.* [49] reported a method for the sustainable and economic transformation of waste product lignin (by-product of paper and pulp industry) into heteroatom doped AC used for electrodes of supercapacitor and $CO_2$ capture applications. The synthesis process involves carbonization and chemical activation. The synthesised AC contains 2.5 to 5.6 wt.% nitrogen and 54 wt.% oxygen in its final structure. It possesses a high surface area 1788 to 2957 $m^2$/g, capacitance of 372 F/g and excellent cyclic stability over 30,000 cycles in 1 KOH. Razmjooei *et al.* [50] developed AC from the most available human waste, urine. It started with the removal of mineral salt from urine carbon (URC), which makes it more porous followed by heteroatom doping (N, S and P). The combined effect of surface properties and porous structure makes it feasible for energy storage applications. It exhibits 1040.5 $m^2$/g surface area, good conductivity and heteroatom doping of N, S and P exhibiting capacitance of 166 F/g at 0.5 A/g in a three-electrode method of testing.

*Graphene derived carbon*

In perspective, graphene is the most promising electrode material for various energy storage applications in particular supercapacitors due to a high electrical conductivity, high SSA, and excellent mechanical strength [51]. Its porous structure also facilitates charge transport in the supercapacitor. These exceptional properties of graphene make it a suitable candidate for the supercapacitor electrode.

The SSA of graphene is highly tuneable according to the required supercapacitor electrode for energy storage applications. Also, the presence of highly movable free pi ($\pi$) electrons on its orbital are responsible for the exceptionally high electrical conductivity [51]. Furthermore, the electrical behaviour of graphene can be improved through functionalization [52] and heteroatom doping [53]. Torabi *et al.* [54] synthesised nanocomposite electrodes constituting porous graphene nanoribbons (PGNRs) and carbon black (CB). This PGNRs/CB electrode has a large specific area (1062.5 m$^2$/g) and capacity of 223.0 F/g at 1.0 A/g current density. Supercapacitor electrodes were developed by intercalating copolymer Pluronic F127 between the layers of reduced graphene oxide (rGO) sheets. The intercalation of copolymer increases the surface area and pore-volume that results in the enhancement of surface wettability and improves its electrochemical performance [55]. In another study, hydrazine reduced graphene hydrogel (GH-Hz) electrode possesses high electrical conductivity, with a specific capacity of 220 F/g at 1 A/g and a high retention capability (around 74%) at a high current density (100 A/g) [56]. Xu *et al.* reported an efficient way to prepare holey graphene oxide through a scalable defect-etching strategy that creates numerous nanopores across the GO plane. Further reduction of holey graphene oxide using H$_2$O results in three-dimensional hierarchical porous holey graphene hydrogel with significantly enhanced ion transport and surface area [57].

## Supplementary Note 2: Data source

Table S1. Information on the literature sources analysed to implement the dataset.

| S. No. | Year | Title | Figure /Table | Reference |
|---|---|---|---|---|
| 1 | 2010 | Microstructure and electrochemical double-layer capacitance of carbon electrodes prepared by zinc chloride activation of sugar cane bagasse | Figure 7a | https://doi.org/10.1016/j.jpowsour.2009.08.048 |
| 2 | 2011 | Preparation of Highly Conductive Graphene Hydrogels for Fabricating Supercapacitors with High Rate Capability | Figure 5a & 5b | https://doi.org/10.1021/jp204036a |
| 3 | 2011 | Hierarchical porous carbon obtained from animal bone and evaluation in electric double-layer capacitors | Figure 5 | https://doi.org/10.1016/j.carbon.2010.10.025 |
| 4 | 2011 | Preparation of capacitor's electrode from sunflower seed shell | Figure 2a & 2b | https://doi.org/10.1016/j.biortech.2010.08.110 |
| 5 | 2012 | Activated carbons from KOH-activation of argan (Argania spinosa) seed shells as supercapacitor electrodes | Figure 6 | doi:10.1016/j.biortech.2012.02.010 |
| 6 | 2012 | Preparation of activated carbon from cotton stalk | Figure 6 | https://doi.org/10.1007/s10008-012-1946-6 |
| 7 | 2012 | Carbonized Chicken Eggshell Membranes with 3D Architectures as High-Performance Electrode Materials for Supercapacitors | Figure 3d | https://doi.org/10.1002/aenm.201100548 |
| 8 | 2013 | High-Performance Asymmetric Supercapacitor Based on Nanoarchitectured Polyaniline/Graphene/Carbon Nanotube and Activated Graphene Electrodes | Figure 4d | https://doi.org/10.1021/am4028235 |
| 9 | 2013 | Rice husk-derived porous carbons with high capacitance by ZnCl2 activation for supercapacitors | Figure 5b | http://dx.doi.org/10.1016/j.electacta.2013.05.050 |
| 10 | 2013 | From coconut shell to porous graphene-like nanosheets for high-power supercapacitors† | Figure 7c | https://doi.org/10.1039/C3TA10897J |
| 11 | 2013 | Tunable N-doped or dual N, S-doped activated hydrothermal carbons derived from human hair and glucose for supercapacitor applications | Figure 4d | https://doi.org/10.1016/j.electacta.2013.06.065 |
| 12 | 2013 | Human hair-derived carbon flakes for electrochemical supercapacitors | Figure 5d | https://doi.org/10.1039/C3EE43111H |
| 13 | 2013 | Preparation of activated carbon hollow fibers from ramie at low | Figure 3c | https://doi.org/10.1016/j.biortech.2013.09.026 |
| 14 | 2013 | Efficient preparation of biomass-based mesoporous carbons for supercapacitors with both high energy density and high power density | Figure 3a | https://doi.org/10.1016/j.jpowsour.2013.03.174 |
| 15 | 2013 | Nitrogen-Doped Porous Graphitic Carbon as an Excellent Electrode Material for Advanced Supercapacitors | Figure 8a | https://doi.org/10.1002/chem.201303345 |
| 16 | 2014 | Hierarchical porous and N-doped carbon nanotubes derived from polyaniline for electrode materials in supercapacitors | Figure 6d | https://doi.org/10.1039/C4TA01465K |

| 17 | 2014 | Freestanding 3D mesoporous graphene oxide for high performance energy storage applications | Figure 5c | https://doi.org/10.1039/C4RA08519A |
| --- | --- | --- | --- | --- |
| 18 | 2014 | Colossal pseudocapacitance in a high functionality–high surface area carbon anode doubles the energy of an asymmetric supercapacitor | Figure 3d | https://doi.org/10.1039/C3EE43979H |
| 19 | 2014 | Importance of open, heteroatom-decorated edges in chemically doped-graphene for supercapacitor applications | Figure 8a | https://doi.org/10.1039/C4TA00936C |
| 20 | 2014 | Shape-controlled porous nanocarbons for high performance supercapacitors | Figure 6d | https://doi.org/10.1039/C3TA15245F |
| 21 | 2014 | A novel route for preparation of high-performance porous carbons from hydrochars by KOH activation | Table 2 | https://doi.org/10.1016/j.colsurfa.2014.01.013 |
| 22 | 2014 | A high-performance carbon derived from corn stover via microwave and slow pyrolysis for supercapacitors | Table 1 | https://doi.org/10.1016/j.jaap.2014.07.010 |
| 23 | 2014 | Surfactant-modified chemically reduced graphene | Figure 9b | https://doi.org/10.1039/C4RA03826F |
| 24 | 2014 | Oriented and Interlinked Porous Carbon Nanosheets with an | Figure 4c | https://doi.org/10.1021/cm503845q |
| 25 | 2014 | Superior capacitive performance of active carbons deSuperior capacitive performance of active carbons derived from Enteromorpha prolifera | | http://dx.doi.org/10.1016/j.electacta.2014.04.101 |
| 26 | 2014 | Hierarchical nitrogen-doped porous carbon with high surface area derived from endothelium corneum gigeriae galli for high-performance supercapacitor | Figure 5d | https://doi.org/10.1016/j.electacta.2014.03.015 |
| 27 | 2014 | Direct Synthesis of Highly Porous Interconnected Carbon Nanosheets and Their Application as High-Performance Supercapacitors | Figure 5b | https://doi.org/10.1021/nn501124h |
| 28 | 2015 | Converting biowaste corncob residue into high value added porous carbon for supercapacitor electrodes | | https://doi.org/10.1016/j.biortech.2015.04.005 |
| 29 | 2015 | Nitrogen-doped hierarchical porous carbon materials prepared from meta-aminophenol formaldehyde resin for supercapacitor with high rate performance | Figure 5/Table 1 & 2 | http://dx.doi.org/10.1016/j.electacta.2014.11.075 |
| 30 | 2015 | Nitrogen-doped porous carbon derived from biomass waste for high-performance supercapacitor | Figure 2e | http://dx.doi.org/10.1016/j.biortech.2015.07.100 |
| 31 | 2015 | Promising biomass-based activated carbons derived from willow catkins for high performance supercapacitors | Figure 7d | http://dx.doi.org/10.1016/j.electacta.2015.03.048 |
| 32 | 2015 | Promising Nitrogen-Rich Porous Carbons Derived from One-Step Calcium Chloride Activation of Biomass-Based Waste for High Performance Supercapacitors | Figure 5g & s11 | https://doi.org/10.1021/acssuschemeng.5b00926 |
| 33 | 2015 | High capacitive performance of exfoliated biochar nanosheets from biomass waste corn cob | Figure 6d | https://doi.org/10.1021/acssuschemeng.5b00926 |
| 34 | 2015 | Ultrahigh Surface Area Three-Dimensional Porous Graphitic Carbon | Figure 6c | https://doi.org/10.1021/acscentsci.5b00149 |

| | | from Conjugated Polymeric Molecular Framework | | |
|---|---|---|---|---|
| 35 | 2015 | Ultrahigh volumetric capacitance and cyclic stability of fluorine and nitrogen co-doped carbon microspheres. | | https://doi.org/10.1038/ncomms9503 |
| 36 | 2015 | Activated porous carbon prepared from paulownia flower for high performance supercapacitor electrodes | Figure 3d | http://dx.doi.org/10.1016/j.electacta.2014.12.169 |
| 37 | 2015 | Hierarchically porous carbon by activation of shiitake mushroom for capacitive energy storage | Figure 5c | http://dx.doi.org/10.1016/j.carbon.2015.05.056 |
| 38 | 2015 | Large scale production of biomass-derived nitrogen-doped porous carbon materials for supercapacitors | Figure 6b | http://dx.doi.org/10.1016/j.electacta.2015.04.082 |
| 39 | 2015 | Impregnation assisted synthesis of 3D nitrogen-doped porous carbon with high capacitance | Figure 7d | https://doi.org/10.1016/j.carbon.2015.07.058 |
| 40 | 2015 | High performance electrode materials for electric double-layer capacitors based on biomass-derived activated carbons | Figure 5g | https://doi.org/10.1016/j.electacta.2015.05.080 |
| 41 | 2015 | Solution Processable Holey Graphene Oxide and Its Derived Macrostructures for High-Performance Supercapacitors | Figure 3d, 4h, 5b, 6c, & 6f | https://doi.org/10.1021/acs.nanolett.5b01212 |
| 42 | 2015 | Facile self-templating large scale preparation of biomass-derived 3D hierarchical porous carbon for advanced supercapacitors | Figure 5e | https://doi.org/10.1039/C5TA04721H |
| 43 | 2015 | Impregnation assisted synthesis of 3D nitrogen-doped porous carbon with high capacitance | Figure 7d | https://doi.org/10.1016/j.carbon.2015.07.058 |
| 44 | 2015 | Electrochemical properties of carbon from oil palm kernel shell for high performance supercapacitors | Figure 8d | https://doi.org/10.1016/j.electacta.2015.05.163 |
| 45 | 2015 | Nitrogen, oxygen and phosphorus decorated porous carbons derived from shrimp shells for supercapacitors | Figure 5c | https://doi.org/10.1016/j.electacta.2015.07.094 |
| 49 | 2016 | Construction of nitrogen-doped porous carbon buildings using interconnected ultra-small carbon nanosheets for ultra-high rate supercapacitors | Figure 4d | https://doi.org/10.1039/C6TA02570F |
| 50 | 2016 | Renewable Graphene-Like Nitrogen-Doped Carbon Nanosheets as Supercapacitor Electrodes with Integrated High Energy-Power Property | Figure 3c | https://doi.org/10.1039/C6TA02828D |
| 51 | 2016 | A melamine-assisted chemical blowing synthesis of N-doped activated carbon sheets for supercapacitor application (Activated carbon) | Figure 5c | http://dx.doi.org/10.1016/j.jpowsour.2016.04.069 |
| 52 | 2016 | Hierarchically Porous N-Doped Carbon Nanosheets Derived From Grapefruit Peels for High-Performance Supercapacitors | Figure 5c | http://dx.doi.org/10.1002/slct.201600133 |
| 53 | 2016 | Facile Synthesis of Three-Dimensional Heteroatom-Doped and Hierarchical Egg-Box-Like Carbons Derived from Moringa oleifera Branches for High-Performance Supercapacitors | Figure 5e | https://doi.org/10.1021/acsami.6b10893 |
| 54 | 2016 | A shiitake-derived nitrogen/oxygen/phosphorus co-doped | Figure 4c | https://doi.org/10.1039/C6RA13689C |

| # | Year | Title | Figure/Table | DOI/Link |
|---|------|-------|--------------|----------|
| | | carbon framework with hierarchical trimodal porosity for high-performance electrochemical capacitors | | |
| 55 | 2016 | A Two-Step Etching Route to Ultrathin Carbon Nanosheets for High Performance Electrical Double Layer Capacitors | Figure 5c | https://doi.org/10.1039/C6NR02155G |
| 56 | 2016 | Heteroatom-Doped Porous Carbon Nanosheets: General Preparation and Enhanced Capacitive Properties | Figure 5c | https://doi.org/10.1002/chem.201602922 |
| 57 | 2016 | Effect of pristine graphene incorporation on charge storage mechanism of three-dimensional graphene oxide: superior energy and power density retention | Figure 5b | 10.1038/srep31555 |
| 58 | 2016 | KOH-Activated Porous Carbons Derived from Chestnut Shell with Superior Capacitive Performance | Figure 7d | https://doi.org/10.1002/cjoc.201600320 |
| 59 | 2016 | Multi-heteroatom self-doped porous carbon derived from swim bladders for large capacitance supercapacitors | Figure 7c | https://doi.org/10.1039/C6TA06337C |
| 60 | 2016 | Promising porous carbons derived from lotus seedpods with outstanding supercapacitance performance | Table 1/Figure 4b | http://dx.doi.org/10.1016/j.electacta.2016.05.020 |
| 61 | 2016 | Preparation and application of capacitive carbon from bamboo shells by one step molten carbonates carbonization | Figure 6c | http://dx.doi.org/10.1016/j.ijhydene.2016.05.083 |
| 62 | 2016 | Biomass-Swelling Assisted Synthesis of Hierarchical Porous Carbon Fibers for Supercapacitor Electrodes | Figure 4c & 5c | https://doi.org/10.1021/acsami.5b11558 |
| 63 | 2016 | Hierarchical structured carbon derived from bagasse wastes: A simple and efficient synthesis route and its improved electrochemical | Figure 5e | https://doi.org/10.1016/j.jpowsour.2015.10.063 |
| 64 | 2016 | Hierarchical Porous Carbon Microtubes Derived from Willow Catkins for Supercapacitor Application | Figure 6e | https://doi.org/10.1039/C5TA09043A |
| 65 | 2016 | Nitrogen-doped interconnected carbon nanosheets from pomelo mesocarps for high performance supercapacitors | Figure 5d | http://dx.doi.org/10.1016/j.electacta.2015.12.195 |
| 66 | 2016 | Nitrogen-doped mesoporous carbons for high performance supercapacitors | Table 4 | http://dx.doi.org/10.1016/j.apsusc.2016.04.064 |
| 67 | 2016 | Pumpkin-Derived Porous Carbon for Supercapacitors with High Performance | Figure 4e | https://doi.org/10.1002/asia.201600303 |
| 68 | 2016 | Popcorn-Derived Porous Carbon for Energy Storage and CO2 Capture | Figure 3b | https://doi.org/10.1021/acs.langmuir.6b01953 |
| 69 | 2016 | A new route for the fabrication of corn starch-based porous carbon as electrochemical supercapacitor electrode material | | https://doi.org/10.1016/j.colsurfa.2016.05.049 |
| 70 | 2016 | Preparation of activated carbon from willow leaves and evaluation in electric double-layer- capacitors | Figure 3c | https://doi.org/10.3390/molecules25184255 |
| 71 | 2016 | Microporous carbon from a biological waste-stiff silkworm for capacitive energy storage | | https://doi.org/10.1016/j.electacta.2016.10.120 |
| 72 | 2017 | Extremely high-rate aqueous supercapacitor fabricated using doped carbon nanoflakes with large surface | Figure 3d | https://doi.org/10.1007/s12274-017-1486-6 |

| | | area and mesopores at near-commercial mass loading | | |
|---|---|---|---|---|
| 73 | 2017 | High performance aqueous supercapacitor based on highly nitrogen doped carbon nanospheres with unimodal mesoporosity . | Figure 4e & 5d | http://dx.doi.org/10.1016/j.jpowsour.2016.10.086 |
| 74 | 2017 | Designed formation of hollow particle-based nitrogen-doped carbon nanofibers for high-performance supercapacitors | Figure 5c | https://doi.org/10.1039/C7EE00488E |
| 75 | 2017 | Highly Doped Carbon Nanobelts with Ultrahigh Nitrogen Content as High-Performance Supercapacitor Materials | Figure 4c | https://doi.org/10.1002/smll.201700834 |
| 76 | 2017 | Multiscale Pore Network Boosts Capacitance of Carbon Electrodes for Ultrafast Charging | Figure 3c | https://doi.org/10.1021/acs.nanolett.7b00533 |
| 77 | 2017 | Porous carbon derived from ailanthus altissima with unique honeycomb-like microstructure for high-performance | Figure 4e | https://doi.org/10.1039/C7NJ01127J |
| 78 | 2017 | Preparation of highly porous carbon through activation of NH4Cl induced hydrothermal microsphere derivation of glucose | Figure 6c | https://doi.org/10.1039/C6RA26141H |
| 79 | 2017 | Hierarchical Hybrids Integrated by Dual Polypyrrole-Based Porous Carbons for Enhanced Capacitive Performance | Figure 5d | https://doi.org/10.1002/chem.201702544 |
| 80 | 2017 | Supercapacitor electrode materials with hierarchically structured pores from carbonization of MWCNTs | Table S3 | https://doi.org/10.1039/C6NR08987A |
| 81 | 2017 | Engineered Fabrication of Hierarchical Frameworks with Tuned Pore Structure and N,O-Co-Doping for High-Performance Supercapacitors | Figure 3c | https://doi.org/10.1021/acsami.7b09801 |
| 82 | 2017 | Enzymatic hydrolysis lignin derived hierarchical porous carbon for supercapacitors in ionic liquids with high power and energy densities | Figure 3c | https://doi.org/10.1039/C7GC00506G |
| 83 | 2017 | Hierarchical nitrogen-doped porous carbon derived from lecithin for high-performance supercapacitors | Figure 5f | https://doi.org/10.1039/C7RA07984B |
| 84 | 2017 | Superior supercapacitive performance of hollow activated carbon nanomesh with hierarchical structure derived from poplar catkins | Figure 7e | http://dx.doi.org/10.1016/j.jpowsour.2017.07.021 |
| 85 | 2017 | Biomass based nitrogen-doped structure-tunable versatile porous carbon material | Figure S9a | https://doi.org/10.1039/C7TA02113E |
| 86 | 2017 | Template-free synthesis of N-doped carbon with pillared-layered pores as bifunctional materials for supercapacitor and environmental applications | Figure 5c | http://dx.doi.org/10.1016/j.carbon.2017.03.027 |
| 87 | 2017 | Promising nitrogen-doped porous nanosheets carbon for supercapacitors | Figure 6d | https://doi.org/10.1007/s11581-016-1897-5 |
| 88 | 2017 | Electrochemical Studies on Corncob Derived Activated Porous Carbon for Supercapacitors Application in Aqueous and Non-aqueous Electrolytes | Figure 6 | https://doi.org/10.1016/j.electacta.2017.01.095 |
| 89 | 2017 | Preparation of high performance supercapacitor materials by fast pyrolysis of corn gluten meal waste | Figure 4c & 5h | https://doi.org/10.1039/C7SE00029D |

| # | Year | Title | Figure | DOI |
|---|------|-------|--------|-----|
| 90 | 2017 | Enhanced electrochemical performance of straw-based porous carbon fibers for supercapacitor | Figure 6c | https://doi.org/10.1007/s10008-017-3689-x |
| 91 | 2017 | Porous 3D Few-Layer Graphene-like Carbon for Ultrahigh-Power Supercapacitors with Well-Defined Structure–Performance Relationship | Figure 3d & 5d | https://doi.org/10.1002/adma.201604569 |
| 92 | 2017 | Fish gill-derived activated carbon for supercapacitor application | Figure 8d | https://doi.org/10.1016/j.jallcom.2016.10.013 |
| 93 | 2017 | An activated carbon derived from tobacco waste for use as a supercapacitor electrode material | Figure 7d | https://doi.org/10.1016/S1872-5805(17)60140-9 |
| 94 | 2017 | Flute type micropores activated carbon from cotton stalk for high performance supercapacitors | Figure 7a | 10.1016/j.jpowsour.2017.05.054 |
| 95 | 2017 | N-doped porous reduced graphene oxide as an efficient electrode material for high performance flexible solid-state supercapacitor | Figure 5e | https://doi.org/10.1016/j.apmt.2016.10.002 |
| 96 | 2018 | One-pot synthesis of nitrogen-doped ordered mesoporous carbon spheres for high-rate and long-cycle life supercapacitors | Figure 7c | https://doi.org/10.1016/j.carbon.2017.10.084 |
| 97 | 2018 | High surface area carbon materials derived from corn stalk core as electrode for supercapacitor | | https://doi.org/10.1016/j.diamond.2018.06.018 |
| 98 | 2018 | Three-dimensional porous activated carbon derived from loofah sponge biomass for supercapacitor applications | Figure 8e | https://doi.org/10.1016/j.apsusc.2017.11.249 |
| 99 | 2018 | Activated biomass carbon made from bamboo as electrode material for supercapacitors | Figure 5c | https://doi.org/10.1016/j.materresbull.2018.03.006 |
| 100 | 2018 | Supercapacitor Electrode Based on Activated Carbon Wool Felt | Figure 9b | https://doi.org/10.3390/c4020024 |
| 101 | 2018 | Waste Biomass Based-Activated Carbons Derived from Soybean Pods as Electrode Materials for High-Performance Supercapacitors | Figure 4c | https://doi.org/10.1002/slct.201800609 |
| 102 | 2018 | A high performance nitrogen-doped porous activated carbon for supercapacitor derived from pueraria | Figure 6c | https://doi.org/10.1016/j.jallcom.2018.02.078 |
| 103 | 2018 | Activated carbons from agricultural waste solvothermally doped with sulphur as electrodes for supercapacitors | Figure 5b | https://doi.org/10.1016/j.cej.2017.11.141 |
| 104 | 2018 | Hierarchical porous carbon prepared from biomass through a facile method for supercapacitor applications | Figure 5b | https://doi.org/10.1016/j.jcis.2018.06.076 |
| 105 | 2018 | N-enriched multilayered porous carbon derived from natural casings for high-performance supercapacitors | Figure 5d | https://doi.org/10.1016/j.apsusc.2018.03.100 |
| 106 | 2018 | Activated carbon derived from harmful aquatic plant for high stable supercapacitors | Figure 3g | https://doi.org/10.1016/j.cplett.2017.11.031 |
| 107 | 2018 | High performance porous graphene nanoribbons electrodes synthesized via hydrogen plasma and modified by Pt-Ru nanoclusters for charge storage and methanol oxidation | Figure 3d | https://doi.org/10.1016/j.electacta.2018.09.082 |
| 108 | 2018 | Sustainable activated carbons from dead ginkgo leaves for supercapacitor electrode active materials | Figure 5f | https://doi.org/10.1016/j.ces.2018.02.004 |
| 109 | 2018 | Tailoring Biomass-Derived Carbon for High-Performance | Figure 5c | https://doi.org/10.1039/C7TA09608A |
| 110 | 2019 | Sakura-based activated carbon preparation and its performance in supercapacitor applications | Figure 7g | https://doi.org/10.1039/C8RA09685F |
| 111 | 2019 | Enhancement of the electrochemical properties of commercial coconut shell-based activated carbon by H2O dielectric barrier discharge plasma | Figure 8d | http://dx.doi.org/10.1098/rsos.180872 |
| 112 | 2019 | Highly Porous Willow Wood-Derived Activated Carbon for High-Performance Supercapacitor Electrodes | Figure 6c | https://doi.org/10.1021/acsomega.9b01977 |
| 113 | 2019 | Oxygen- and Nitrogen-Enriched Honeycomb-Like Porous Carbon from Laminaria japonica with Excellent Supercapacitor Performance in Aqueous Solution | Figure 8a | https://doi.org/10.1021/acssuschemeng.9b01448 |
| 114 | 2019 | A sustainable approach to produce activated carbons from pecan nutshell waste for environmentally friendly supercapacitors | Figure 7c | https://doi.org/10.1016/j.carbon.2019.04.017 |
| 115 | 2019 | N, S co-doped porous carbons from natural Juncus effuses for high performance supercapacitors | Figure 3e | https://doi.org/10.1016/j.diamond.2019.107577 |
| 116 | 2019 | Robust hierarchically interconnected porous carbons derived from discarded Rhus typhina fruits for ultrahigh capacitive performance supercapacitors | Figure 5b | https://doi.org/10.1016/j.jpowsour.2018.12.064 |
| 117 | 2019 | Nitrogen self-doped porous carbon with layered structure derived from porcine bladders for high-performance supercapacitors | Figure 5c | https://doi.org/10.1016/j.jcis.2019.02.024 |
| 118 | 2019 | Nitrogen-doped microporous carbon derived from a biomass waste metasequoia cone for electrochemical capacitors | Figure 5c | https://doi.org/10.1016/j.jallcom.2019.04.237 |
| 119 | 2019 | Low-cost, high-performance supercapacitor based on activated carbon Activated carbon derived from pitaya peel for supercapacitor applications with high capacitance performance | Figure 5a | https://doi.org/10.1016/j.jcis.2018.11.103 |
| 120 | 2020 | Activated carbon derived from pitaya peel for supercapacitor applications with high capacitance performance | | https://doi.org/10.1016/j.matlet.2020.127339 |
| 121 | 2020 | S-doped activated mesoporous carbon derived from the Borassus flabellifer flower as active electrodes for supercapacitors | Table S3 | https://doi.org/10.1016/j.matchemphys.2019.122151 |
| 122 | 2020 | Advanced porous hierarchical activated carbon derived from agricultural wastes toward high performance supercapacitors | Figure 5e | https://doi.org/10.1016/j.jallcom.2019.153111 |
| 123 | 2020 | Activated carbons prepared by indirect and direct CO2 activation of lignocellulosic biomass for supercapacitor electrodes | Figure 8a & 8b | https://doi.org/10.1016/j.renene.2020.03.111 |

| # | Year | Title | Figure | DOI |
|---|------|-------|--------|-----|
| 124 | 2020 | A new method of synthesizing hemicellulose-derived porous activated lignocellulosic biomass for supercapacitor electrodes carbon for high-performance supercapacitors | | https://doi.org/10.1016/j.micromeso.2019.109707 |
| 125 | 2020 | The performance of sulphur doped activated carbon supercapacitors prepared from waste tea | | https://doi.org/10.1080/09593330.2019.1575480 |
| 126 | 2020 | The use of activated carbon from coffee endocarp for the manufacture of supercapacitors | | https://doi.org/10.1007/s10854-020-03123-1 |
| 127 | 2020 | Microporous carbon from malva nut for supercapacitors: Effects of primary carbonizations on structures and performances | Figure 5e | https://doi.org/10.1016/j.diamond.2020.107816 |
| 128 | 2020 | Seaweed-derived KOH activated biocarbon for electrocatalytic oxygen reduction and supercapacitor applications | Figure 7b | https://doi.org/10.1007/s10934-020-00871-7 |
| 129 | 2020 | Soybean-waste-derived activated porous carbons for electrochemical double-layer supercapacitors: Effects of processing parameters | Figure 6e | https://doi.org/10.1016/j.est.2019.101070 |
| 130 | 2020 | Boosting the supercapacitor performances of activated carbon with carbon nanomaterials | Figure 4j | https://doi.org/10.1016/j.jpowsour.2019.227678 |
| 131 | 2020 | Nano-porous carbon materials derived from different biomasses for high performance supercapacitors | Figure 7 | https://doi.org/10.1016/j.ceramint.2019.11.031 |
| 132 | 2020 | An ultrasonic-assisted synthesis of rice-straw based porous carbon with high performance symmetric supercapacitors | Figure 6d | https://doi.org/10.1039/C9RA08537H |
| 133 | 2020 | Hydrangea-like N/O codoped porous carbons for high-energy supercapacitors | Figure 6f | https://doi.org/10.1016/j.cej.2020.124208 |
| 134 | 2020 | Low-cost and advanced symmetry supercapacitors based on three-dimensional tea waste of porous carbon nanosheets | Figure 5a | https://doi.org/10.1080/10667857.2020.1714902 |
| 135 | 2020 | Areca nut–derived porous carbons for supercapacitor and CO2 capture applications | Figure 8 | https://doi.org/10.1007/s11581-019-03261-5 |
| 136 | 2020 | Hierarchical porous carbon electrode materials for supercapacitor developed from wheat straw cellulosic foam | Figure 5e | https://doi.org/10.1016/j.renene.2019.11.150 |
| 137 | 2020 | Heteroatoms-doped hierarchical porous carbon derived from chitin for flexible all-solid-state symmetric supercapacitors | Figure 5d | https://doi.org/10.1016/j.cej.2019.123263 |
| 138 | 2020 | Walnut shell-derived hierarchical porous carbon with high performances for electrocatalytic hydrogen evolution and symmetry supercapacitors | Figure 4e | https://doi.org/10.1016/j.ijhydene.2019.10.159 |
| 139 | 2020 | Facile preparation of N-O codoped hierarchically porous carbon from alginate particles for high performance supercapacitor | Figure 6e | https://doi.org/10.1016/j.jcis.2019.12.027 |
| 140 | 2020 | O/N-co-doped hierarchically porous carbon from carboxymethyl cellulose ammonium for high performance supercapacitors | Figure 6e | https://doi.org/10.1007/s10853-020-04515-8 |
| 141 | 2019 | Hierarchical porous carbon microrods derived from albizia flowers for high performance supercapacitors | Figure 6e | https://doi.org/10.1016/j.carbon.2019.02.072 |
| 142 | 2020 | Scalable green synthesis of hierarchically porous carbon microspheres by spray pyrolysis for high-performance supercapacitors | Figure 8f | https://doi.org/10.1016/j.cej.2019.122805 |
| 143 | 2020 | Hierarchically porous carbon derived from the activation of waste chestnut shells by potassium bicarbonate (KHCO3) for high-performance supercapacitor electrode | Figure 9 | https://doi.org/10.1002/er.4970 |
| 144 | 2020 | Synthesis of porous carbon nanostructure formation from peel waste for low cost flexible electrode fabrication towards energy storage applications | Table 2 | https://doi.org/10.1016/j.est.2020.101735 |
| 145 | 2020 | Nitrogen-doped Oxygen-rich Activated Carbon Derived from Longan Shell for Supercapacitors | Figure 7e | 10.20964/2020.03.18 |
| 146 | 2020 | Mesopore-rich carbon flakes derived from lotus leaves and its ultrahigh performance for supercapacitors | Figure 4f | https://doi.org/10.1016/j.electacta.2019.135481 |
| 147 | 2020 | An Ultra-microporous Carbon Material Boosting Integrated Capacitance for Cellulose-Based Supercapacitors | Figure 3C | https ://doi.org/10.1007/s4082 0-020-0393-7) |

**Supplementary Note 3: Details on the regression models**

*Ordinary least square (OLS) regression*

OLS is one of the most common regression models, where the unknown parameters of linear regression are estimated by lessening the sum of the squares of the differences between the target responses of the sample dataset and the value foreseen by a linear function of explanatory variables [58]. A linear regression can be described as:

$$Y = \beta_0 + \beta_1 X_1 + \beta_2 X_2 + \cdots + \beta_i X_i + \epsilon, \tag{S1}$$

where $Y$ is the dependent variable, $X_i$ is the explanatory variable, $\beta_i$ is the coefficient, and $\epsilon$ is the random error term.

*Support Vector Regression (SVR) model*

SVR is a well-established supervised machine learning approach for predicting discrete values. SVR operates on the same principle as Support Vector Machine (SVM). The primary principle of SVR is to determine the best fit line. An optimal hyperplane defines SVM as a discriminative classifier, whereas – in SVR – the best fit line is the hyperplane with the most point. Support vectors are the results of ideal hyperplanes, which classify unseen datasets that support hyperplanes [59, 60]. The hyperplane in a two-dimensional (2D) region is a line separating into two segments wherein each segment is placed on either side. For instance, multiple line data classification can be done with two distinct datasets (*i.e.,* green and red) and used to propose an affirmative interpretation (see Figure S1). However, selecting an optimal hyperplane is not an easy job, as it should not be noise sensitive, and the generalization of datasets should be accurate [61]. Pertinently, SVM is used to determine the optimized hyperplane that provides considerable minimum distance to the trained dataset. SVR attempts to minimize the difference between the real and predicted values by fitting the best line under a certain threshold value. The distance between the hyperplane and the boundary line is the threshold value (see Figure S1). In mathematical notation, for a 2D space, a line can be used to distinguish linearly separable data. The line can be represented as

$$y = ax + b. \tag{S2}$$

By renaming $x$ with $x_1$ and $y$ with $x_2$, the equation is modified as

$$ax_1 - x_2 + b = 0. \tag{S3}$$

If we substitute $X = (x_1, x_2)$ and $w = (a, -1)$, we get the following:

$$wX + b = 0, \tag{S4}$$

which is called the equation of the hyperplane and refers to the SVM.

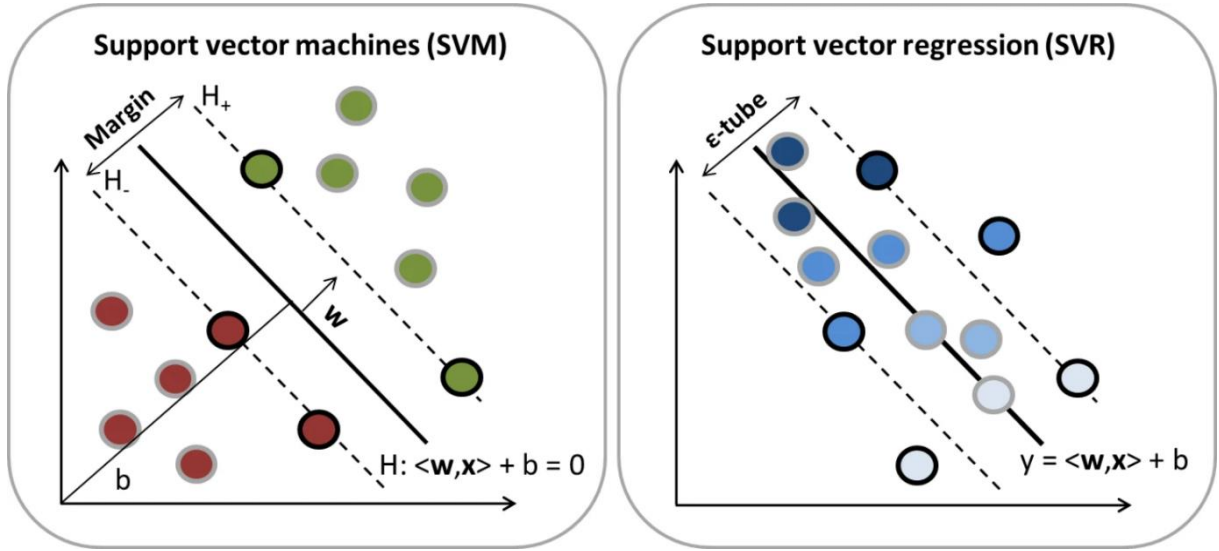

Figure S1: Data classification using (a) SVM and (b) SVR [62].

Consider, the decision boundaries are at any distance say 'ε' from the hyperplane. So, these are the lines that we draw at a distance '+ε' and '-ε' from the hyperplane. Then the equations of decision boundary become:

$$wX + b = +\varepsilon, \quad (S5)$$
$$wX + b = -\varepsilon. \quad (S6)$$

Thus, any hyperplane that satisfies our SVR should satisfy:

$$-\varepsilon < wX + b < +\varepsilon. \quad (S7)$$

The key goal here is to choose a decision boundary that has 'ε' distance from the initial hyperplane and contains data points closest to the hyperplane.

*Decision tree (DT)*
DT constructs the regression or classification models based on the data features in the tree's configuration. In a tree, every node is related to the property of a data feature. Moreover, it either predicts the target value (regression) or predict the target class (classification). The closer the nodes in a tree are, the greater their influence [63]. Some benefits of the DT include:
- It is easy and simple to understand, analyse, and intercept.
- It is capable of handling both categorical and numerical data.

*Random forest (RF)*
 RF is an ensemble learning technique that can perform both regression and classification tasks utilizing the multiple decision trees. During training, the algorithm generates a large number of decision trees using a probabilistic scheme [64]; every tree is trained on a bootstrapped sample of the original training data and finds a randomly selected subset of the input variables to determine a split (for each node). Every tree in the RF makes its own individual prediction or casts a unit vote for the most popular class at input *x*. These predictions are then averaged in case of regression or the majority vote determines the output in case of classification [64]. The core concept is to use numerous decision trees to determine the final output rather than depending on individual decision trees.

*Extreme Gradient Boosting (XGBoost) model*

XGBoost is one application of gradient boosting machines (GBMs) mainly designed for speed and performance. GBM is the most effective algorithms for supervised learning. In supervised learning, various features in the training data are utilized to predict the target values. XGBoost applies the classification and regression trees (CART) algorithms to a known dataset and categorises the data accordingly [65]. For a dataset consisting of *n* number of samples and *m* number of features, $\mathbb{D} = \{(x_i, y_i)\}(|\mathbb{D}| = n, x_i \in R^m, y_i \in R)$, the expression of an XGBoost algorithm, the total number of CART trees is as follows:

$$\hat{y}_i = \sum_{k=1}^{K} f_k(x_i), f_k \in \mathbf{F}, \tag{S8}$$

where $\mathbf{F} = \{f(x) = w_{q(x)}\}(q: R^m \to T, w \in R^T)$ is the CART trees space, and $q(x)$ corresponds to an input $x$ to a leaf node of a CART tree. The symbols $w$ and $T$ represent the weight of the node and sum of the leaves in a tree, respectively. As a result, XGBoost calculates a final score by adding up all the weights from each CART tree. The learning goal is to determine the appropriate weights and splitting threshold for each tree node to reduce model complexity. The total loss function of XGBoost is defined as squared loss plus a regularization term:

$$\mathcal{L} = \sum_i l(\hat{y}_i, y_i) + \sum_i \Omega(f_x) = \sum_i (\hat{y}_i - y_i)^2 + \sum_k \left(\gamma T_k + \frac{1}{2}\lambda \|w_k\|^2\right). \tag{S9}$$

By contrast, RFs reduce this loss function by dividing features based on the most significant Gini information gain and by randomly assembling CART trees. XGBoost converts the loss function into a new scoring function that can be used to choose the best threshold [66]:

$$\tilde{\mathcal{L}}^{(t)}(q) = -\frac{1}{2}\sum_{j=1}^{T} \frac{\left(\sum_{i \in I_j} g_i\right)^2}{\sum_{i \in I_j} h_i + \lambda} + \gamma T, \tag{S10}$$

where $\tilde{\mathcal{L}}^{(t)}(q)$ is the second-order approximation of the loss function at the *t*-th iteration, and $g_i$ and $h_i$ are the first and second-order loss gradient on the *i*-th data, respectively. The instance set of a specific leaf node *j* is $I_j$. As a result, XGBoost can reduce loss iteratively and get better results than other ensemble algorithms.

**Supplementary Note 4: Details on the Artificial Neural Network model**

For the sake of comprehensiveness, we also developed an artificial neural network (ANN) to predict the capacitance of carbon-based supercapacitors. Several features were selected such as specific surface area (SSA), pore size (PS), pore volume (PV), $I_D/I_G$ ratio, potential window (PW), current density (I), oxygen, nitrogen, and sulphur content in the electrode. Based on the database and the selected features, the ANN was built upon three layers: an input layer, thirteen hidden layer and an output layer. The dataset was divided into training and testing set. For each node of the proposed ANN, the *ReLU* function was applied as the activation function, being the cost function as the mean absolute error. After 250 epochs trainings, the capacitance could be predicted from the designed ANN model. From the regression analysis, the correlation coefficient ($R^2$) achieved is 0.72, which indicates a fair prediction capability of this model. The other error metrics obtained (*RMSE*, *b'* and *MAPE*) are also indicated in Table S2: overall, ANN shows a slightly lower prediction accuracy as compared to RF and XGBoost models, at least for the considered conditions.

Table S2. Performance analysis of the different ML models.

| Model | $R^2$ | RMSE | b' | MAPE |
|---|---|---|---|---|
| ANN | 0.72 | 47.3 | 1.06 | 26.32 |
| RF | 0.75 | 43.96 | 0.98 | 27.09 |
| XGBoost | 0.79 | 40.27 | 0.95 | 30.08 |

**Supplementary Note 5: Influence of the carbon material percentage on the specific capacitance**

We also investigated the impact of carbon material percentage on specific capacitance by refining our database to include the carbon material percentage (C) in the electrode material. Among the 4538 data entries of our clean database, information on carbon material percentage was available for 3117 of them and has been included. We then used an XGBoost model on this refined database to determine the effect of carbon material percentage on specific capacitance. Figure S2 (a) illustrates the correlation between actual and predicted specific capacitance for the dataset considering a 6M KOH electrolyte *viz* $R^2 = 0.80$, RMSE = 37.09, b' = 0.97, and MAPE = 25.42. Additionally, we performed a feature analysis on the 6M KOH electrolyte dataset, as shown in Figure S2 (b), which revealed that the specific surface area (SSA), heteroatom doping (N%), and pore size (PS) were still the significant factors influencing specific capacitance, coherently with Figure 6 (b).

To further enhance the regression performance, we refined the datasets based on a specific electrolyte (6M KOH) and testing methods (two and three electrode). The effect of this refinement on the data analysis is shown in Figures S2 (c) and (e), where most data points were positioned near the diagonal line, indicating a strong correlation between actual and predicted specific capacitance values. Moreover, the accuracy of the regression was confirmed by the improved statistics of the XGBoost model fitting for both the two-electrode method ($R^2 = 0.90$, RMSE = 25.26, b' = 0.96, and MAPE = 18.23) and the three-electrode method ($R^2 = 0.91$, RMSE = 23.25, b' = 1.007, and MAPE = 8.39). We conducted feature analyses on supercapacitors with a 6M KOH electrolyte, as depicted in Figures S2 (d) and (f). Our findings revealed that PS, SSA, and defects were the major contributors to the capacitive performance in the two-electrode testing method, while PV, SSA, and PS were the major contributors in the three-electrode method.

It is important to note that the percentage of carbon weight is a significant factor in determining the specific capacitance; however, since our database only includes carbon-based material, we were unable to distinguish it in our analysis. Consequently, we found that other parameters appear to be more prominent.

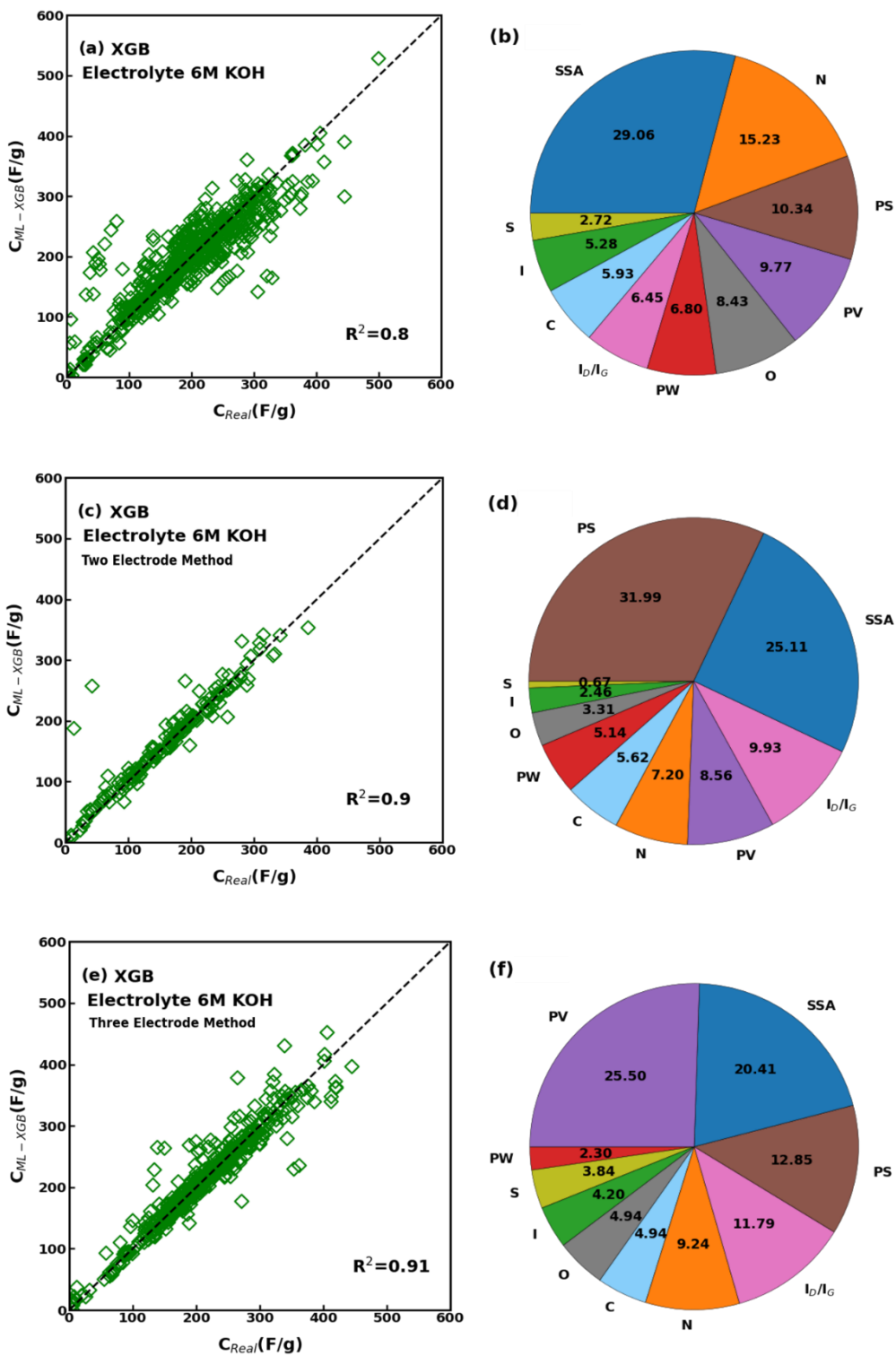

Figure S2: (a) Comparison between the predicted and actual specific capacitance and (b) feature analysis for the subset of data having 6M KOH electrolyte. (c) Comparison between the predicted and actual specific capacitance and (d) feature analysis for the subset of data having 6M KOH electrolyte and measured by two-electrode method. (e) Comparison between the predicted and actual specific capacitance and (f) feature analysis for the subset of data having 6M KOH electrolyte and measured by three-electrode method. These analyses were carried out also considering the carbon material percentage (C) in the electrode.

**Supplementary Note 6: Details of the hyperparameters of the ML models in this study**

| Hyperparameters | DT | SVR | RF | XGBoost |
|---|---|---|---|---|
| Gamma | - | 10 | - | 0.3 |
| Max_depth | 10 | - | 16 | 3 |
| Max_features | Auto | - | - | - |
| Min_samples_split | 5 | - | - | - |
| N_estimators | - | - | 1000 | 500 |
| Min_child_weight | - | - | - | 4 |
| C | - | 10 | - | - |
| Kernel | - | RBF | - | - |
| Epsilon | - | 0.1 | - | - |

Table S3. Hyperparameters of the different ML models used in this study for the Figure 4.

| Cases | Gamma | Max_depth | Min_child_weight | N_estimators |
|---|---|---|---|---|
| Three Electrode | 0.5 | 3 | 5 | 500 |
| Two Electrode | 0.4 | 3 | 4 | 500 |
| 6M KOH | 0.3 | 3 | 5 | 300 |
| 1M $H_2SO_4$ | 0.3 | 3 | 5 | 200 |
| Three electrode & 6M KOH | 0.5 | 4 | 4 | 500 |
| Two electrode & 6M KOH | 0.5 | 4 | 4 | 500 |
| AC | 0.3 | 2 | 4 | 500 |
| HPC | 0.3 | 2 | 5 | 100 |
| HA | 0.3 | 2 | 4 | 300 |

Table S4. Hyperparameters of the XGBoost model used in this study for the Figures 5, 6, 7, and 8 respectively.